\documentclass[conference,compsoc]{IEEEtran}
\ifCLASSOPTIONcompsoc
\else
\fi
\ifCLASSINFOpdf
\else
\fi
\usepackage{cite} 
\usepackage[backref]{hyperref} 
\usepackage[pdftex]{graphicx}
\usepackage[lined,linesnumbered,ruled,norelsize]{algorithm2e} 
\usepackage{algorithmic}
\usepackage{array}
\usepackage{amsthm,amsmath,amssymb,amsfonts}
\usepackage{color} 
\usepackage{subfigure}
\usepackage{bm} 
\usepackage{comment} 
\usepackage{picins} 

\makeatletter
\renewcommand{\maketag@@@}[1]{\hbox{\m@th\normalsize\normalfont#1}}%
\makeatother

\begin{document}
%
\title{Online Learning for Failure-aware Edge Backup of \\ Service Function Chains with the Minimum Latency}

\author{\IEEEauthorblockN{Chen Wang}
\IEEEauthorblockA{School of Computer\\Science and Technology\\
Shandong University, China\\
Email: chenw@mail.sdu.edu.cn}
\and
\IEEEauthorblockN{Qin Hu}
\IEEEauthorblockA{Department of Computer\\and Information Science\\
IUPUI, USA\\
Email: qinhu@iu.edu}
\and
\IEEEauthorblockN{Dongxiao Yu}
\IEEEauthorblockA{School of Computer\\Science and Technology\\
Shandong University, China\\
Email: dxyu@sdu.edu.cn}
\and
\IEEEauthorblockN{Xiuzhen Cheng}
\IEEEauthorblockA{School of Computer\\Science and Technology\\
Shandong University, China\\
Email: xzcheng@sdu.edu.cn}}


%


\maketitle

\begin{abstract}
Virtual network functions (VNFs) have been widely deployed in mobile edge computing (MEC) to flexibly and efficiently serve end users running resource-intensive applications, which can be further serialized to form service function chains (SFCs), providing customized networking services. To ensure the availability of SFCs, it turns out to be effective to place redundant SFC backups at the edge for quickly recovering from any failures. The existing research largely overlooks the influences of \textit{SFC popularity, backup completeness,} and \textit{failure rate} on the optimal deployment of SFC backups on edge servers. In this paper, we comprehensively consider from the perspectives of both the end users and edge system to backup SFCs for providing popular services with the lowest latency.
To overcome the challenges resulted from unknown SFC popularity and failure rate, as well as the known system parameter constraints, we take advantage of the online bandit learning technique to cope with the uncertainty issue. Combining the \textit{Prim}-inspired method with the greedy strategy, we propose a \textit{Real-Time Selection and Deployment} (\textit{RTSD}) algorithm. 
Extensive simulation experiments are conducted to demonstrate the superiority of our proposed algorithms.
\end{abstract}


%
\IEEEpeerreviewmaketitle

\section{Introduction}
An explosive amount of service requests are generated at the network edge due to the increasing number of smart devices connecting to the Internet to run various real-time applications, demanding higher bandwidth and lower latency. 
To serve end users more efficiently, the emerging mobile edge computing (MEC) paradigm aims to provide a variety of computing and networking services to users in a physically-close manner \cite{mao2017survey,xiao2019edge}, which can greatly reduce the response delay for users' requests and the possibility of network congestion, comprehensively improving the user experience.

In recent, network virtualization has been proposed to efficiently manage networking resources, which has also been widely implemented at MEC through network function virtualization (NFV). In particular, to disengage network functions from specific hardware, virtual network functions (VNFs) are utilized in NFV to facilitate scalable and flexible operations for service providers \cite{erfanian2017network,van2020vnf,peuster2019introducing,chiosi2014network}. Further, multiple VNFs can be connected sequentially to compose a service function chain (SFC), which aims at providing specialized networking services more efficiently \cite{fei2018adaptive,zhang2019raba}.
However, due to the vulnerability of VNFs in deployment \cite{taleb2016service}, it is challenging to ensure the availability of SFCs because any failed VNF component can invalidate the entire service chain \cite{shang2018partial,shang2020reducing,fan2018framework,shang2018placement}. 

To deal with this challenge, some existing studies focus on the optimal deployment of VNFs at the beginning of the configuration process \cite{sang2017provably,kuo2018deploying,mijumbi2016connectionist,cziva2018dynamic,jin2020latency,zhu2018edgeplace}, while others rely on the idea of preparing backups of SFCs so that they can quickly recover from any unexpected failures of VNFs \cite{fan2017availability,kanizo2017optimizing,dinh2018efficient,wang2021proactive}. 
In practice, SFC backups can be placed on both the central cloud server and distributed edge servers. Considering that routing to the central cloud is generally costly in operation and time consumption, researchers suggest backing up SFCs at the network edge \cite{dinh2018efficient,shang2020reducing}, where the limitation of edge resource has to be rigorously studied. 
However, as a critical reflection of the users' preferences, \textit{popularity} of SFCs has long been overlooked in the design of SFC backup scheme at the edge, which can act as an index in selecting the appropriate set of SFC backups given the limited resource available at the edge. 
Though a recent work \cite{wang2021proactive} takes into account the user demands in deploying VNF backups at the edge, the \textit{completeness} of SFCs is not carefully treated with scattered VNF backups on edge servers. More importantly, the \textit{failure rate} of each SFC has rarely been considered in the backup deployment schemes, which also refers to a significant property of SFCs in availability guarantee.

In this paper, we comprehensively consider from the perspectives of both the end users and edge system, incorporating the impacts of SFC popularity, completeness, and failure rate on the deployment of SFC backups at the edge.
Nevertheless, there exist two aspects of challenges from both the \textit{unknown} and \textit{known} factors. To be specific, the unknown challenge is due to the fact that either the popularity or failure rates of SFCs cannot be known as a priori to better select the set of SFCs being backed up on edge servers. No matter for the service requests from end users or the happening of SFC failure, it can be highly dynamic and uncertain for the system design. While the other challenge from the known factors concerns the systematic parameters, including the resource constraint of edge servers to reasonably place SFC backups, the latency minimization of all SFC backups for providing efficient responses to users, and the binary deployment variable to place any SFC backup either at the edge or the cloud. 
The combination of uncertainty in the unknown challenge and confirmed constraints in the known challenge results in the difficulty to appropriately deploy SFC backups at the edge for fundamentally and efficiently enhancing the service availability.

To resolve the above challenges, we utilize the online bandit learning to have an estimation of unknown information and gradually approach the real values according to the instantly realistic feedback through both exploration and exploitation. With the learning results of unknown parameters, we optimize the link latency of all SFCs based on a variant of Prim algorithm solving the Minimum Spanning Tree (MST) problem, based on which a greedy strategy based VNF backup deployment algorithm is designed to finally place relevant VNFs composing selected SFCs on edge servers.

In summary, our major contributions in this work can be listed as follows:
\begin{itemize}
    \item Concerning the request preferences of end users and the changing environment, we model the SFC backup placement at the edge as an optimization problem, aiming to provide popular services with the lowest latency, where the resource constraint of edge servers and backup completeness of SFCs are strictly prescribed.
    \item To solve the proposed optimization problem with the unknown information, we resort to the combinatorial multi-armed bandit (CMAB) problem to learn the popularity and failure rate of SFCs in an online manner. Combined with the Prim-inspired solution and greedy strategy, we propose a \textit{Real-Time Selection and Deployment} (\textit{RTSD}) algorithm to achieve near-optimal placement of SFC backups at the edge.
    \item To verify the superiority of our proposed algorithm, we design two benchmark solutions for comparison. Moreover, the performance of our scheme is testified for varying end users and edge parameters.
\end{itemize}

The remaining of this paper is organized as follows. Section \ref{sec:related} investigates the most related work about SFC backup. Section \ref{sec:problem} presents the system model and problem formulation of deploying SFC backups at the edge, which is addressed by our proposed RTSD algorithm in Section \ref{sec:alg}. Section \ref{sec:experiment} displays experimental results and Section \ref{sec:conclusion} concludes our paper.


\section{Related Work}\label{sec:related}
The advancement of NFV technology has effectively solved the problems of traditional networking system, 
where the successful deployment of VNF is the key to realize the virtualization \cite{erfanian2017network,van2020vnf,peuster2019introducing,chiosi2014network}. To provide more sophisticated services to end users, connecting multiple VNFs to form the SFC becomes prevailing \cite{fei2018adaptive,jin2020latency,fan2017availability,zhang2019raba}. 
However, how to ensure the availability of SFCs has long been a challenge \cite{shang2018partial,kanizo2017optimizing,dinh2018efficient,shang2020reducing,fan2017availability,fan2018framework,shang2018placement,zhu2018edgeplace,zhang2019raba}. To solve this problem, using redundant technology to create backups of SFCs turns out to be an effective approach. 
Fan~\textit{et al.} \cite{fan2017availability} adopt an availability-aware SFC mapping method with off-site redundancy while reducing resources consumption. Kanizo~\textit{et al.} \cite{kanizo2017optimizing} introduce a certain number of backups which share the available data and resources of VNFs to achieve the purpose of coping with SFC failure. Dinh~\textit{et al.} \cite{dinh2018efficient} propose a cost-efficient deployment scheme based on measuring the improvement potential of VNFs and various techniques for redundancy placement to guarantee the availability of SFCs.

Given a large number of SFC backups, how to deploy them in the network system becomes a critical problem as placing all backups on the central cloud server can bring huge routing cost and service delay. The emergence of edge computing provides us with some new solution. 
When it comes to placing SFC backups at the edge, most of the existing studies focus on the deployment and routing issues, aiming to achieve less resource consumption since the available resources of edge servers are constrained \cite{jin2020latency,cziva2018dynamic,zhu2018edgeplace,dinh2018efficient,shang2020reducing,wang2021proactive}. 
Among them, Jin~\textit{et al.} \cite{jin2020latency} try to deploy SFCs at the edge with the goal of consuming the lowest edge resource and bandwidth, which has to satisfy the constraint of total latency of each SFC as well. Zhu~\textit{et al.} \cite{zhu2018edgeplace} consider both the available resources and deployment costs, aiming at minimizing the cost of deploying SFC at the edge. However, the above studies mainly focus on placing the original VNFs at the edge, so the service availability cannot be guaranteed in the event of VNF failure. To solve this issue, Wang~\textit{et al.} \cite{wang2021proactive} select qualified VNFs to be backed up at the edge thinking about the users' demands, but the atomicity and completeness of SFCs, as well as the VNF failure, are not considered at all, which cannot comprehensively solve the availability issue of SFCs.

To fundamentally enhance the SFC availability, we focus on deploying SFC backups at the edge with the minimum service delay under the resource constraint of edge environment, considering both the users' real-time service requests and potential failures of VNFs, where each SFC can only be entirely backed up at the edge or the cloud. 
Specifically, we formulate two optimization problems and propose two bandit learning based solutions to achieve the greatest reward for deploying SFC backups at the edge. 

\section{System Model}\label{sec:problem}
\begin{figure}[htbp]
\centering
\centerline{\includegraphics[width=0.48\textwidth]{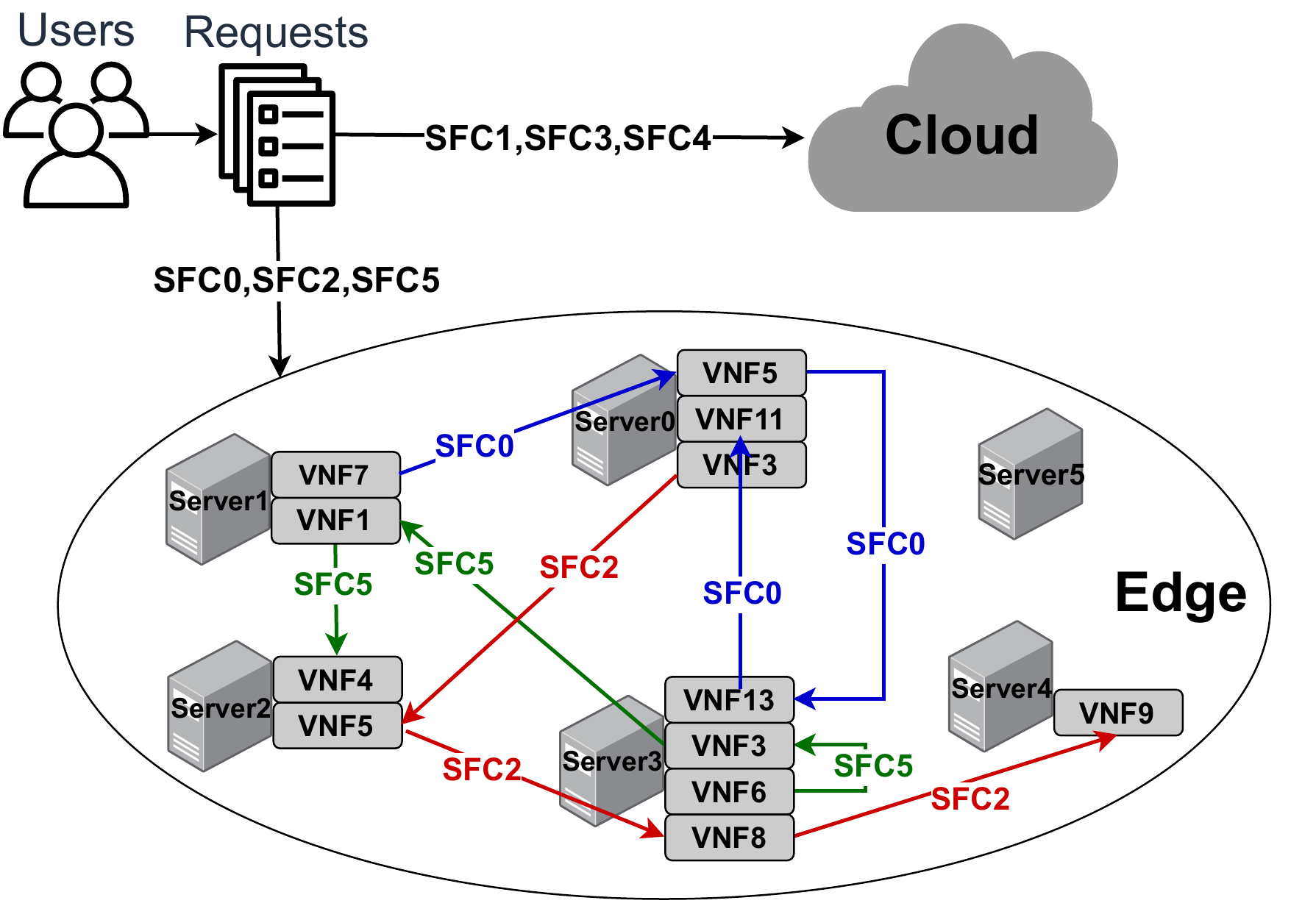}}
\caption{An example of SFC backup deployment at the edge.}
\label{fig}
\end{figure}

In the context of mobile edge computing, the entire edge environment can be regarded as a network model, which can be modeled as an undirected connected graph $\mathcal G(\mathcal N, \mathcal E)$, with the set of nodes $\mathcal N$ representing all servers at the edge of network and the set of edges $\mathcal E$ denoting the network links among servers. Fig. \ref{fig} shows a brief demonstration of entire system, which contains a central cloud server and an edge network consisting of multiple servers. In this figure, three SFCs, i.e., SFC0 (blue), SFC2 (red), and SFC5 (green), are backed up at the edge; 
while the backups of the remaining three SFCs are deployed at the cloud. For reference, we summarize main notations used in this paper in Table \ref{tab:notations}.

\begin{table}[]
    \centering
    \footnotesize
    \caption{Notations.}
    \label{tab:notations}
    \begin{tabular}{|p{0.7cm}<{\centering}|p{6.8cm}|}
    \hline
    Symbol & Explanation \\
    \hline
    $\mathcal{N}$   &The set of edge servers $\mathcal N=\{1,2,\cdots,n,\cdots,N\}$ \\
    \hline
    $M_n$   &The available resource of edge server $n$\\
    \hline
    $\mathcal{K}$   &The set of users $\mathcal K=\{1,2,\cdots,k,\cdots,K\}$\\
    \hline
    $\mathcal{F}$   &The set of SFCs $\mathcal F=\{1,2,\cdots,f,\cdots,F\}$\\
    \hline
    $\mathcal{I}$   &The set of VNFs $\mathcal I=\{1,2,\cdots,i,\cdots,I\}$\\
    \hline
    $D_i$   &Resource requirements for backing up VNF $i$ at the edge \\
    \hline
    $I_f$   &The set of VNFs composing an SFC $f$\\
    \hline
    $Y_{i,f}$   &Binary variable showing whether VNF $i$ is involved in SFC $f$\\
    \hline
    $Q_{k,f}(t)$  &Binary variable showing whether user $k$ has a service request for SFC $f$ during time slot $t$\\
    \hline
    $Q_f(t)$   &Total requests for SFC $f$ during time slot $t$\\
    \hline
    $q_f(t)$   &The expected popularity of SFC $f$ during time slot $t$\\
    \hline
    $X_f(t)$   &Binary variable showing whether SFC $f$ is placed at the edge during time slot $t$\\
    \hline
    $P_{i,n}^j(t)$   &Binary variable showing whether VNF $i$ is placed on edge server $n$ during time slot $t$\\
    \hline
    $l(u,v)$   &The latency of link between edge server $u$ and $v$\\
    \hline
    $\mathcal{L}$   &The set of link latency\\
    \hline
    $\phi_{f(u,v)}^j$   &Count variable showing whether the $j$th link of chain $f$ deployed at the edge passes through link $(u,v)$\\
    \hline
    $L_f(t)$   &The total delay of SFC $f$ during time slot $t$\\
    \hline
    $U_f(t)$   &The failure rate of SFC $f$ during time slot $t$\\
    \hline
    $V_i(t)$   &The failure rate of VNF $i$ during time slot $t$\\
    \hline
    $v_i(t)$   &The expected failure rate of VNF $i$ during time slot $t$\\
    \hline
    $R_f(t)$   &Service hit reward brought by deploying the backup of SFC $f$ at the edge during time slot $t$\\
    \hline
    \end{tabular}
\end{table}

\subsection{Basic Parameters}
Assuming that the collection of edge servers is represented by $\mathcal N=\{1,2,\cdots,n,\cdots,N\}$ with $N$ denoting the total number of edge servers, where the available storage resource of any edge server $n$ to backup VNFs is defined as $M_n$. We denote the set of users by $\mathcal K=\{1,2,\cdots,k,\cdots,K\}$ with $K$ being the number of all users. User $k$ will issue multiple service requests during time slot $t$, and each service request has to be responded by the corresponding SFC. All SFCs are included in the set $\mathcal F=\{1,2,\cdots,f,\cdots,F\}$ and all VNFs are represented by the set $\mathcal I=\{1,2,\cdots,i,\cdots,I\}$, in which $F$ and $I$ are respectively the numbers of SFCs and VNFs; and the resource requirement of backing up each VNF $i$ on the edge server is defined as $D_i$. 
Besides, the VNFs composing an SFC $f$ is denoted by the set $I_f \subseteq \mathcal I$ and there are $|I_f|$ VNFs in SFC $f$, which includes $|I_f|-1$ links. Define a binary variable $Y_{i,f}\in \{0,1\}$ to indicate whether VNF $i$ is involved in SFC $f$ or not, i.e., $i\in I_f $ or not. If VNF $i$ is a member of SFC $f$, we have $Y_{i,f}=1$; otherwise, $Y_{i,f}=0$. 

\subsection{Popularity of SFCs}\label{subsec:demand}
Since our scheme selects SFCs to backup at the edge from the perspective of users' requests, it is necessary to obtain the service requests from users in each time slot for calculating the popularity of each SFC.

We define a binary variable $Q_{k,f}(t)\in \{0,1\}$ to denote whether user $k$ has a service request for SFC $f$ in time slot $t$, where $Q_{k,f}(t)=1$ if the service request sent by user $k$ in time slot $t$ is realized through SFC $f$ and $Q_{k,f}(t)=0$ when SFC $f$ is not requested by user $k$. 
Then the total requests of all users for SFC $f$ in time slot $t$, denoted by $Q_f(t)$, can be calculated as
\begin{equation}\label{eq:Q}
Q_f(t)=\sum_{k\in\mathcal{K}}Q_{k,f}(t).
\end{equation}

It is clear that $Q_f(t)$ is a discrete random variable in the set of 
$\{0,1,2,\cdots,k,\cdots,K\}$, which has an average mathematical expectation, denoted by $q_f(t)$, as follows,
\begin{equation}\label{eq:q}
q_f(t) = \mathbb{E}[Q_f(t)].
\end{equation}

In this article, we use this expected value $q_f(t)$ to express the popularity of SFC $f$ during each time slot $t$, that is, the total expected expectation of service requests for SFC $f$ sent by users in each time slot. We assume that the popularity of each SFC remains stable within the time slot range. 

\subsection{Backup Deployment Variables}\label{backup deployment}
In order to reflect whether a certain SFC is selected to be backed up at the edge, a binary variable $X_f(t) \in \{0,1\}$ is used to represent the backup placement decision for SFC $f$ during each time slot $t$. If it is decided to deploy the backup of SFC $f$ at the edge in time slot $t$, $X_f(t)=1$; otherwise, $X_f(t)=0$.

Since SFCs are formed by linking required VNFs in sequence, the SFC backup deployment will eventually be implemented as the deployment of multiple VNFs. To indicate the backup placement decision of a certain VNF at the edge, we introduce a binary variable $P_{i,n}(t) \in \{0,1\}$ to represent whether the backup of VNF $i$ is placed on edge server $n$ in time slot $t$. And we have $P_{i,n}(t)=1$ if VNF $i$ is stored on server $n$ and $P_{i,n}(t)=0$ if not. 

To ensure that a certain SFC can provide complete functions at the edge, it is required that only when all VNFs in this chain are deployed on edge servers, can this SFC be considered to be backed up at the edge network successfully; otherwise, it will be placed in the cloud. According to this requirement, we can get the relationship between $X_f(t)$ and $P_{i,n}(t)$ as follows: 
\begin{equation}\label{eq:X and P}
X_f(t)=1- \min \{ 1, \sum_{i\in I_f}(1-P_{i,n}(t)) \}.
\end{equation}

\subsection{Resource Constraint}
To back up SFCs at the edge successfully, the resource constraint of edge servers cannot be ignored. In detail, the total resource requirement of all backups of VNFs placed on server $n$ should not exceed the available resource of this server, i.e., $M_n$, which can be expressed as follows: 
\begin{equation}\label{eq:resource constraint}
\sum_{i\in\mathcal{I}}D_i*P_{i,n}(t)\leq M_n.
\end{equation}

\subsection{Link Latency of SFCs}
We assume that the latency of link $(u,v)$ between the edge server $u$ and edge server $v$ is relatively stable and denote it as $l(u,v)$, all of which constitute a collection $\mathcal{L} = \{l(u,v),(u,v)\in \mathcal{E}\}$. For each SFC $f$, we use a count variable $\phi_{f(u,v)}^j$ to indicate whether the $j$th link of this service chain $f$ deployed at the edge passes through the link $(u,v)$. 
The value of $\phi_{f(u,v)}^j$ is set to 1 when the $j$th link of $f$ passes through $(u,v)$, and 0 otherwise. Then the total delay of SFC $f$ deployed at the edge network in time slot $t$, defined as $L_f(t)$, can be calculated by: 
\begin{equation}\label{eq:latency}
L_f(t)=\sum_{j=1}^{|I_f|-1}\phi_{f(u,v)}^j*l(u,v).
\end{equation}

\subsection{Failure Rate Calculation}
We all know that the operation of an SFC is based on the successful functioning of all VNFs composing this SFC. The failure of any one will lead to the failure of the entire SFC. Therefore, we are supposed to further consider improving the availability of SFC backups with the awareness that every SFC has a certain possibility to come into a failure. When this happens unfortunately, no matter how popular the SFC is and how low the latency it achieves, this SFC cannot successfully provide services to users and they won't get any benefits. We introduce $U_f(t)$ to represent the failure rate of SFC $f$ in time slot $t$, and we have the success rate as $1-U_f(t)$, which is jointly affected by the failure rates of the contained VNFs. Therefore, the problem is further transformed into obtaining the failure rate of each VNF.  

We assume that the real-time failure rate of VNF $i$ during time slot $t$ is $V_i(t)$, which is a random variable within the range $[0, 1]$. After choosing one VNF to work each time, observe whether it fails (or whether it provides services successfully). By pulling this arm many times, it can be found that the distribution tends to stabilize. In other words, for VNF $i$, the probability of failing to realize the function has a fixed expected value, denoted by $v_i(t)$, from which we can get:
\begin{equation}\label{eq:v}
v_i(t) = \mathbb{E}[V_i(t)].
\end{equation}
Then the failure rate of VNF $i$ is $v_i(t)$ and the success rate is $1-v_i(t)$. 

Since the failure of any VNF contained in each SFC will lead to the unavailability of the entire chain, we define that the failure rate of this SFC is the largest failure rate of all VNFs in the chain, that is:
\begin{equation}\label{eq:SFC_failure}
U_f(t)= \max_{i \in I_f} v_i(t).
\end{equation}

In the above formula, we consider the worst case, that is, we consider the maximum probability of failure for a certain SFC as its own failure rate. This approach is to ensure the availability of our scheme in the event of a failure, making it more reliable. The specific value method needs to be determined by the service provider based on actual practice. But all results can be applied to the algorithms we proposed in this paper. 

\subsection{Backup Reward}
If an SFC corresponding to a user's request is backed up at the edge, 
there is no need to forward this request to the cloud. This clearly avoids additional traffic, reducing the propagation delay and users' access cost, which can be regarded as the improvement of users' service hit reward. 
To compare the hit reward brought by different backup schemes at the edge, we use $R_f(t)$ to represent the service hit reward brought by SFC $f$ that satisfies the users' service request during time slot $t$, where the value of $R_f(t)$ is influenced by three factors, i.e., the popularity, the link delay and the failure rate of SFC $f$. In detail, deploying SFCs with high popularity $Q_f(t)$ at the edge can greatly reduce the cost of being routed to the cloud, especially for the frequently requested SFCs, which brings a higher service hit reward. Besides, when a certain SFC $f$ is deployed at the edge, the lower the link delay $L_f(t)$, the faster the users' service requests can be satisfied, and the greater service hit reward would be obtained. Finally, the failure of one SFC can invalid the service provision to users, which will make the hit reward become zero. 

Therefore, we define the service hit reward brought by deploying the backup of SFC $f$ at the edge as: 
\begin{equation}\label{eq:reward}
R_{f}(t)=(\omega Q_{f}(t)-\mu L_{f}(t))*X_f(t)*(1-U_f(t)),
\end{equation}
where $\omega, \mu>0$ are constant scalars. Note that in the practical implementation, the relationship between these parameters can be determined by the service provider according to the specific environment and preferences. 

\subsection{Problem Formulation}
According to the above description, our problem is to select the appropriate set of SFCs and deploy them at the edge to maximize time-average service hit reward with the comprehensive consideration of users' needs, SFC composition, and link latency, on the premise of satisfying edge resource constraint. Therefore, we can get the SFC backup problem formulated as: 
\begin{subequations}\label{eq:problem}
\begin{align}
\mathrm{max} \qquad &\frac{1}{T}\sum_{t=0}^{T-1}\sum_{f\in \mathcal{F}}\mathbb{E}[R_f(t)], \label{eq:optimization_goal}\\
\mathrm{s.t.}\qquad &(\ref{eq:X and P})(\ref{eq:resource constraint}),\notag\\
&P_{i,n}(t)\in\{0,1\},\forall n\in\mathcal{N}, \forall i\in\mathcal{I}, t, \label{eq:constraint_1}\\
&X_f(t)\in\{0,1\},\forall f\in\mathcal{F}, t. \label{eq:constraint_2}
\end{align}
\end{subequations}

In the above equations, the optimization goal in \eqref{eq:optimization_goal} is to maximize the time-average service hit reward; constraint \eqref{eq:X and P} describes the relationship between the VNF backup variable $P_{i,n}(t)$ and the SFC backup variable $X_f(t)$; \eqref{eq:resource constraint} is the resource constraint; constraints \eqref{eq:constraint_1} and \eqref{eq:constraint_2} show that SFCs and VNFs are either deployed at the edge or deployed in the cloud. 

\section{Solution of SFC Backup Problem: \textit{RTSD}}\label{sec:alg}
In this section, to solve the SFC backup problem in the edge environment, we first conduct a further analysis in Section \ref{Problem Reformulation}. Subsequently, the \textit{RTSD} solution is proposed in Section \ref{RTSD} to deal with the existing difficulties, and the specific implementation and detailed algorithm designs of each component in our scheme are introduced in Sections \ref{subsec:link latency} to  \ref{subsec:Selection and deployment}. 
\subsection{Problem Reformulation}\label{Problem Reformulation}
From \eqref{eq:optimization_goal}, we can know that the ultimate goal of our SFC backup problem is to maximize the time-average service hit reward. Substituting the definition of reward in \eqref{eq:reward} into the above problem, we can further specify the optimization goal as 
\begin{equation}\label{eq:problem2}
\mathrm{max}~\frac{1}{T}\sum_{t=0}^{T-1}\sum_{f\in \mathcal{F}}\mathbb{E}[(\omega Q_{f}(t)-\mu L_{f}(t))*X_f(t)*(1-U_f(t))].
\end{equation}

It is clear that the deployment strategy $X_f(t)$ of SFC $f$ is independent of the request $Q_f(t)$, the link delay $L_f(t)$ and the failure rate $U_f(t)$. In addition, since the popularity $Q_f(t)$ has an expectation $q_f(t)$ from \eqref{eq:q} and the failure rate of VNF $V_i(t)$ has an expectation $v_i(t)$ from \eqref{eq:v}, we can further transform the above equation into
\begin{footnotesize}
\begin{equation}\label{eq:problem3}
\mathrm{max}~\frac{1}{T}\sum_{t=0}^{T-1}\sum_{f\in \mathcal{F}}\mathbb{E}[X_f(t)]*(\omega q_{f}(t)-\mu \mathbb{E}[L_{f}(t)])*(1-\max_{i \in I_f} v_i(t)).
\end{equation}
\end{footnotesize}

We can see that the value of the reward is affected by three aspects, which is positively proportional to the popularity $q_f(t)$ of the chain $f$ while inversely proportional to the total delay $L_f$ of SFC $f$ deployed at the edge and the maximum failure rate of VNFs in each SFC. Thus, we need to find the optimal strategy to select the popular set of SFCs to deploy on the appropriate edge servers so as to obtain the biggest service hit reward. 

Intuitively, we can decompose the problem of SFC backup selection and deployment at the edge into four steps: 
\begin{enumerate}
    \item For a specific SFC, we back up all involved VNFs on edge servers with limited resources, making the link delay of this chain as low as possible;
    \item Obtaining the popularity $q_f(t)$ of each SFC;
    \item Obtaining the failure rate $v_i(t)$ of each VNF in each SFC;
    \item According to the above results, we can get the maximum service hit reward corresponding to each SFC, and we give priority to the chain with the largest reward so that it will be backed up at the edge based on the minimum delay deployment plan until no backup resources available.
\end{enumerate}

In the above process, the key of completing the selection of SFC backups is to learn the actual values of popularity $q_f(t)$ and failure rate $v_i(t)$. However, in practice, because users' service requests and SFC status change dynamically in each time slot and it is difficult to predict them in advance, the popularity of different chains and the failure rate of each VNF become unknown priori, making it challenging to obtain their values.
\subsection{Real-Time Selection and Deployment}\label{RTSD}
To address the issue mentioned above, inspired by the Upper Confidence Bound (UCB) algorithm, that is, deigned for solving the multi-armed bandit learning problem \cite{auer2002finite}, we can analyze the historical information captured before and estimate the popularity and failure rate under the current time slot combined with real-time users' service requests and surroundings. In this process, we are faced with a challenge that cannot be ignored, that is, how to deal with the trade-off between exploration and exploitation. Among them, the former is to explore potential new arms that may generate higher returns without considering previous experience, which is a radical plan. The advantage is that we can find options with higher reward, but we cannot use the existing high-return arms. The latter develops and uses known arms that can produce high returns based on the content that has been explored to get the best strategy so far. The advantage of this scheme is that it can make full use of existing knowledge, but because of this, it is only limited to the current local optimal scheme and may ignore the better content that has not been explored, thus missing the opportunity to generate higher reward, which is a conservative choice. Therefore, we need to find a balance between two contradictions, so as to maximize the service hit reward at each time slot.

Based on the above ideas, we model our SFC backup problem with unknown popularity and failure information as a multi-armed bandit learning problem, where SFCs can be considered as multiple arms, and the service hit reward obtained by deploying a certain SFC at the edge is the reward brought by pulling an arm. During each time slot, combining historical data, we learn the current popularity of all SFCs according to users' requests and the real-time failure rate of VNFs according to current environment, calculate the reward of deploying each SFC, and then make a selection based on the calculation result to determine which SFC should be placed at the edge in the current time slot to get the highest return. The deployment result will be used as feedback information to participate in the selection and deployment of SFCs in the next time slot.

\begin{figure}[htbp]
\centering
\centerline{\includegraphics[width=0.48\textwidth]{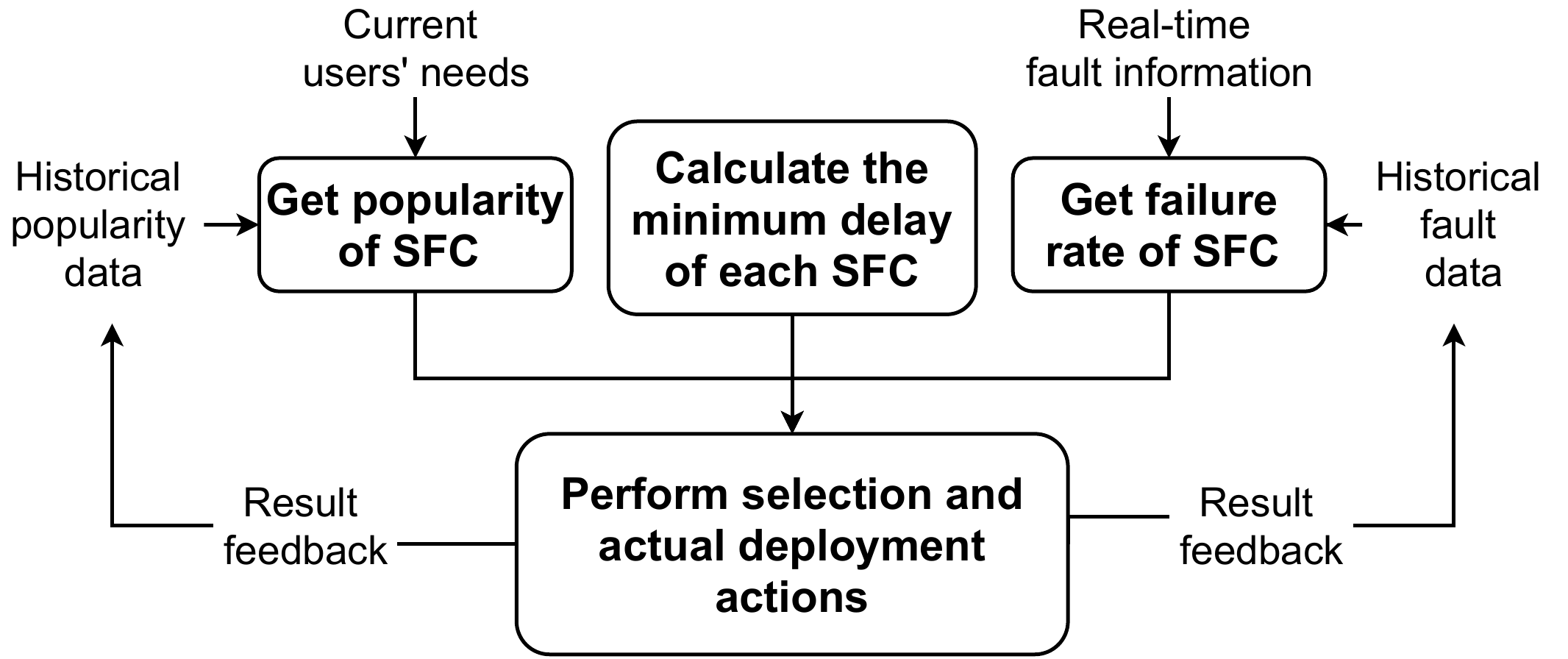}}
\caption{An illustration of \textit{RTSD}.}
\label{fig:RTSD}
\end{figure}

We propose an algorithm named \textit{Real-Time Selection and Deployment} (\textit{RTSD}) based on the above ideas, whose main idea is illustrated by Fig. \ref{fig:RTSD} for a more logical presentation. 

In Algorithm \ref{alg:SFC}, we describe the proposed \textit{RTSD} solution in a formal way, which is implemented with five main parts, namely \textit{Initialization} at the begining and \textit{MinimizeDelay, SFCPlacement, Popularity, FailureRate} in every subsequent time slot. When the whole process begins, we have no historical information about SFCs and VNFs, so the value of each SFC's popularity and VNF's failure rate can only be estimated based on service requests collected from end users. For this reason, we have to set the number of times when each SFC is learned at the edge and the average value of popularity, denoted as $c_{f}(0)$ and $ \bar{q}_{f}(0)$, respectively, as zero, which will participate in learning the estimate of popularity. And the popularity can be considered as the sum of service requests issued by all users for each SFC during time slot $0$, that is,  $\Tilde{q}_{f}(0)=Q_f(0)$. Similarly, we set both $h_{i}(0)$ and $\bar{v}_{i}(0)$ respectively to zero and $\Tilde{v}_{i}(0)$ to the real-time information in time slot $0$, i.e., $V_i(0)$ (Lines 1-6). 

In the second step, our goal is to obtain the minimum link latency of each SFC 
(Lines 9-11). And subsequently, we 
choose the best set of SFCs and deploy the corresponding VNFs on determined edge servers based on the popularity, the failure rate and the minimum delay obtained to get the 
biggest service hit reward (Lines 12-14). The selection and deployment actions will stop when the edge network can no longer accommodate any SFC (Line 8). After the above actions are completed, the SFC backup deployment task of the current time slot has been implemented. Then we will update the historical information according to the deployment result in this time slot. Combined with the users' requests, we get new popularity of each SFC (Lines 16-18). Next we will also reckon the latest failure rate based on historical data (Lines 19-21) and then calculate the minimum delay again, so as to prepare for the SFC selection and deployment task in the next time slot. The above steps will be repeated for each subsequent time slot to select a set of qualified SFCs and properly deploy them on the edge.

The specific implementation of the related sub-algorithms will be elaborated in the following subsections.

\begin{algorithm}\label{alg:SFC}
       \caption{\textit{RTSD}}
       \SetKwBlock{Initialization}{Initialization}{end}
       \SetKwBlock{MinimizeDelay}{MinimizeDelay}{end}
       \SetKwBlock{SFCPlacement}{SFCPlacement}{end}
       \SetKwBlock{Popilarity}{Popilarity}{end}
       \SetKwBlock{FailureRate}{FailureRate}{end}
       \KwIn{ edge server set $\mathcal N$, user set $\mathcal K$, SFC set $\mathcal F$, and VNF set $\mathcal I$}
        \Initialization{
         \ForEach{SFC $f \in \mathcal F$}{
         $c_{f}(0)=0,\bar{q}_{f}(0)=0,\Tilde{q}_{f}(0)=Q_f(0)$\;
         $h_{i}(0)=0,\bar{v}_{i}(0)=0,\Tilde{v}_{i}(0)=V_i(0)$\;
         }
        }
        \ForEach{time slot $t\in \{0,1,\cdots\}$}{
        \While{edge resources are sufficient}{
        \MinimizeDelay{
        \textsc{\textsc{\textsc{GetConsumption}}}($t$)\;
        }
        \SFCPlacement{
        \textsc{\textsc{\textsc{VNFDeployment}}}($t$)\;
        }
        }
        \Popilarity{
        \textsc{\textsc{\textsc{GetPopularity}}}($t$)\;
        }
        \FailureRate{
        \textsc{\textsc{\textsc{GetFailureRate}}}($t$)\;
        }
        }
\end{algorithm}

\subsection{Link Delay of SFC Backups}\label{subsec:link latency}
\subsubsection{Realization idea}

In this part, our task is to arrange multiple VNFs contained in one chain on appropriate edge servers with limited resources. 
The goal is to find the optimal deployment scheme with the lowest total delay consumed by this SFC.

In edge network, there is a certain delay between any two edge servers, the value of which is specified in advance. Our task is to find the connection sequence of a group of available edge servers to make the link delay consumed by connecting them the lowest so that the SFC backed up in this path will be the most efficient. In order to analyze this abstract problem more clearly, we can regard edge servers as nodes in a graph. The delay between servers is materialized into the link with certain weight between two nodes, so that the edge network constitutes a weighted undirected complete graph, where each node has its own capacity (that is, the available resource of each server). Under this specific model, our problem can be seen as \textit{finding a path with the smallest sum of weights that meets the capacity requirement from a complete graph}.
 
Inspired by the problem of Minimum Spanning Tree (MST), we use the idea of Prim algorithm to solve this problem. However, it is worth noting that our problem is different from MST in the following two aspects:
 \begin{itemize}
    \item We all know that the MST problem is to find the smallest set of edges that contains all nodes in the original graph while keeping the graph connected. But in our scenario, we do not need to include all nodes, i.e., edge servers, in the graph, and each node can be selected repeatedly. This is because any certain SFC does not have to travel through all edge servers to realize the specified service;
    \item Our problem has a resource constraint for each edge server, which is not involved in the traditional MST problem. Specifically, each node has a certain amount of resources for backing up SFCs. 
\end{itemize}

In spite of these differences, the same thing is that we all need to achieve the smallest sum of weights for the selected edges. To this end, we learn from the Prim algorithm dealing with the MST problem,  based on which we make appropriate expansions and improvements to form an effective solution for our SFC backup problem. 

\subsubsection{Algorithm implementation}
\begin{algorithm}\label{alg:Placement}
       \caption{\textsc{GetConsumption}($t$)}
       \KwIn{the current time slot $t$, VNF set $\mathcal{I}$, the resource request of each VNF $\mathcal{D}$, edge link set $\mathcal{E}$, link latency set $\mathcal{L}$, the available resources of edge server $\mathcal{M}$}
       \KwOut{the lowest link latency consumed by each SFC \{$L_f(t)$\}}
       Find link $l'$ with the lowest latency\;
       Find server $n'$ with larger resource capacity at both ends of $l'$\;
       $\mathcal{P}\leftarrow{\phi}$\;
       \ForEach{SFC $f\in\mathcal{F\setminus P}$}{
         $\mathbf{CurrentNode}\leftarrow n'$\;
         \ForEach{$i\in I_f$}{
           \While{$\mathbf{CurrentNode}$ can hold current VNF}{
           Place VNF $i$ on $\mathbf{CurrentNode}$\;
           }
           Traverse the link emanating from $\mathbf{CurrentNode}$ in the order of increasing delay\;
           $\mathbf{CurrentNode} \leftarrow $ the edge server that can accommodate VNF $i$ at the other end of the link\;
         }
         \If{all VNF $\in I_f$ are palced on determined servers}
         {
         Calculate the total delay consumed by SFC $f$\;
         }
         $\mathcal{P}=\mathcal{P}\cup \{f\}$\;
       }
\end{algorithm}

In Algorithm \ref{alg:Placement}, we present the step of \textsc{GetConsumption} in detail, where the minimum delay of each SFC and its optimal deployment at the edge are calculated. For a SFC $f$, put the contained VNFs on the determined edge servers in order. Since we need to minimize the total link delay, we give priority to the two servers with the least latency. In addition, a server with larger capacity is easier to accommodate more VNFs, which is more likely to bring zero time consumption for the corresponding links. These two strategies allow us to get the minimum delay of each SFC in the edge environment. Therefore, we first consider placing VNFs on the larger end of the link with the smallest weight (Lines 1-2). Continue to put the VNFs in SFC $f$ on the server until it cannot accommodate the current VNF (Lines 7-9). Learning from the idea of Prim algorithm, we diffuse outward from the current server node. Taking the consumption into account, we traverse the diverging edges of the current node in the order of increasing 
latency until we find a server with available resources that can be fully used to deploy the current VNF (Line 10). Update the current node (Line 11) and repeat the above process until all VNFs in this SFC have been traversed. For this chain, there are two results. First, each VNF in SFC $f$ can be deployed on a certain server (Lines 13-15). At this time, the data chain formed by transmitting data packets in the order of VNFs consumes the minimum of latency. Add delay of each link to get the total latency consumed by this SFC (Line 14). Otherwise, one or several VNFs in this SFC cannot be deployed at the edge. Specifically, no server can meet their resource requirements. In this case, \eqref{eq:X and P} shows that when there are VNFs in one SFC deployed at the cloud, we believe that the backups of the entire chain need to be deployed in the cloud network. At this point, the process of solving the minimum delay of current SFC is over, and we need to continue processing other chains. In order to prevent double calculation, we introduce a set $\mathcal{P}$ to store the traversed SFCs (Line 3), SFCs that have been judged will be put into $\mathcal{P}$ (Line 16), then we can select from the remaining SFCs for calculation (Line 4).

\subsection{Learning Popularity of SFCs}\label{subsec:popularity}
\subsubsection{Realization idea}\label{subsub:popularity}

As analyzed at the beginning of Section \ref{RTSD}, we need to make full use of historical information and combine the current data to estimate uncertain attributes. In order to make full use of the historical service request information that has been obtained, we define a variable $c_f(t)$ to represent the number of time slots when SFC $f$ is learned in the edge network, that is, the time slots when SFC $f$ is selected to be backed up at the edge, which can be calculated as: 
\begin{equation}\label{eq:c_f}
c_f(t)=\sum_{\tau=0}^{t-1}X_{f}(\tau).
\end{equation}

In addition, we define a variable $\bar{q}_{f}(t)$ to record the average value of the learned popularity of SFC $f$ up to time slot $t$, and its formula is described as follows: 
\begin{equation}\label{eq:history_q}
\bar{q}_{f}(t)=\frac{\sum_{\tau=0}^{t-1}Q_f(\tau)}{c_f(t)}.
\end{equation}

During each time slot $t$, the current demand $Q_f(t)$ of each SFC is obtained based on users' service requests, according to which we calculate the average value of learned popularity up to time slot $t$. With reference to the UCB algorithm, we then combine the historical information and current requests to estimate the popularity of SFC, expressed as $\Tilde{q}_{f}(t)$, which is changing continuously. The calculation of $\Tilde{q}_{f}(t)$ is as follows: 
\begin{equation}\label{eq:learned_q}
\Tilde{q}_{f}(t)={\bar{q}_{f}(t)+K\sqrt{\frac{3\log{t}}{2c_{f}(t)}}}.
\end{equation}

According to \eqref{eq:c_f}-\eqref{eq:learned_q}, we can learn the popularity of SFC $f$ in time slot $t$, and further obtain the service hit reward $R_f(t)$ of the chain, which will directly determine the deployment decision $X_f(t)$ in edge environment, i.e., determining which SFC backup to be placed at the edge. After the above actions are completed, we need to update the involved parameter values, i.e., $c_f(t)$ and $\bar{q}_{f}(t)$, based on the decision result so as to continue contributing to the selection and deployment process in next time slot. The corresponding update formulas are defined as follows: 
\begin{equation}\label{eq:update_c}
c_{f}(t+1)=\begin{cases}
c_{f}(t),             &X_f(t)=0,\\
c_{f}(t)+1,           &X_f(t)=1,
\end{cases}
\end{equation}
\begin{equation}\label{eq:update_history}
\bar{q}_{f}(t+1)=\begin{cases}
\bar{q}_{f}(t),          &X_f(t)=0,\\
\frac{c_{f}(t)\bar{q}_{f}(t)+Q_{f}(t+1)}{c_{f}(t+1)},         &X_f(t)=1.
\end{cases}
\end{equation}

\subsubsection{Algorithm implementation}
We design a \textsc{GetPopularity} algorithm to deal with the popularity which is changing dynamically, as shown in Algorithm \ref{alg:popularity}. According to the description in Section \ref{subsub:popularity}, to estimate the popularity of each SFC in the next time slot, we first need to calculate its corresponding parameter information, that is, $c_f(t)$ and $\bar{q}_{f}(t)$, according to the SFC backup placement strategy of the current time slot, $X_f(t)$ (Lines 1-6). The specific parameter update process is defined in \eqref{eq:update_c} and \eqref{eq:update_history}. After mastering the two major parameters involved in estimating popularity, we will implement the online learning step according to \eqref{eq:learned_q} (Lines 7-13). The estimate value of each SFC's popularity is obtained in Line 10 of Algorithm \ref{alg:popularity}.
\begin{algorithm}\label{alg:popularity}
       \caption{\textsc{GetPopularity}($t$)}
       \SetKwBlock{Initialization}{Initialization}{end}
       \SetKwBlock{SFCPlacement}{SFCPlacement}{end}
       \SetKwBlock{OnlineLearning}{OnlineLearning}{end}
       \SetKwBlock{VNFPreDeployment}{VNFPreDeployment}{end}
       \SetKwBlock{GetParameters}{GetParameters}{end}
       \KwIn{the current time slot $t$, the number of learning slots \{$c_{f}(t)$\}, the average value of historical popularity \{$\bar{q}_{f}(t)$\}}
       \KwOut{the popularity of each SFC for next time slot \{$\Tilde{q}_{f}(t+1)$\}}
        \GetParameters{
          \ForEach{SFC $f \in \mathcal F$}{
            $c_{f}(t+1)=\begin{cases}
            c_{f}(t),             &X_f(t)=0\\
            c_{f}(t)+1,           &X_f(t)=1
            \end{cases}$
            
            $\bar{q}_{f}(t+1)=\begin{cases}
            \bar{q}_{f}(t),          &X_f(t)=0\\
            \frac{c_{f}(t)\bar{q}_{f}(t)+Q_{f}(t+1)}{c_{f}(t+1)},         &X_f(t)=1
            \end{cases}$
          }
        }
        \OnlineLearning{
          \ForEach{SFC $f \in \mathcal F$}{
             \If{$c_{f}(t+1)>0$}{
             $\Tilde{q}_{f}(t+1) ={\bar{q}_{f}(t+1)+K\sqrt{\frac{3\log{t}}{2c_{f}(t+1)}}}$
             }
          }
         }
\end{algorithm}


\subsection{Learning Failure Rate of SFCs}\label{subsec:faliure rate}
\subsubsection{Realization idea}\label{subsub:faliure}
Similar to the SFC popularity, the failure rate of any VNF is also unknown a priori, which changes every time slot and cannot be known in advance. Therefore, we also adopt the UCB-based idea to learn its value. We first introduce a variable $h_i(t)$ to record the total number of time slots when VNF $i$ is learned until time slot $t$. In Section \ref{backup deployment}, we use $P_{i,n}(t)$ to reflect the placement decision of VNF $i$, based on which, we can calculate the value of $h_i(t)$:
\begin{equation}\label{eq:h_i}
h_i(t)=\sum_{\tau=0}^{t-1}\sum_{n\in \mathcal{N}}P_{i,n}(\tau).
\end{equation}

Then, the history information denoted by $\bar{v}_{i}(t)$, that is, the average value of all failure rate of this VNF also needs to be obtained according to the following formula:
\begin{equation}\label{eq:history_v}
\bar{v}_{i}(t)=\frac{\sum_{\tau=0}^{t-1}V_i(\tau)}{h_i(t)}.
\end{equation}

Using the above two variable values, we can estimate the value of the current failure rate in time slot $t$ with the help of UCB algorithm. The formula is described as follows:
\begin{equation}\label{eq:learned_v}
\Tilde{v}_{i}(t)={\bar{v}_{i}(t)+K\sqrt{\frac{3\log{t}}{2h_{i}(t)}}}.
\end{equation}

With the consideration of SFC failure, we will get a comprehensive reward by integrating the popularity, link latency and the failure rate obtained above based on \eqref{eq:reward}. According to this value, the deployment decision of each SFC, i.e., $X_f(t)$, can be determined, which will affect the popularity in the next time slot as seen in \eqref{eq:update_c} and \eqref{eq:update_history}. In addition, the deployment decision of VNF, i.e., $P_{i,n}(t)$, can also be found out, which serves as the feedback of the failure rate of next slot. The corresponding updates are as follows: 
\begin{equation}\label{eq:update_h}
h_{i}(t+1)=\begin{cases}
h_{i}(t),             &\sum_{n\in \mathcal{N}}P_{i,n}(t)=0,\\
h_{i}(t)+\sum_{n\in \mathcal{N}}P_{i,n}(t),           &\sum_{n\in \mathcal{N}}P_{i,n}(t)>0,
\end{cases}
\end{equation}

\begin{equation}\label{eq:update_v}
\bar{v}_{i}(t+1)=\begin{cases}
\bar{v}_{i}(t),          &\sum_{n\in \mathcal{N}}P_{i,n}(t)=0,\\
\frac{h_{i}(t)\bar{v}_{i}(t)+V_{i}(t+1)}{h_{i}(t+1)},         &\sum_{n\in \mathcal{N}}P_{i,n}(t)>0.
\end{cases}
\end{equation}

\subsubsection{Algorithm implementation}
Using Algorithm \ref{alg:failure}, we present the calculation process of SFC failure rate. After the initialization is completed in time slot $0$, we make a random selection without learning. In the subsequent time slots, the values of related parameters, i.e., the number of time slots when VNF is learned $h_i(t)$ and the average value of history data $\bar{v}_{i}(t)$, are updated based on the selection result of the previous time slot (Lines 2-5), which will participate in the estimation of the failure rate according to UCB Algorithm (Lines 8-12). Because the smooth function of SFC depends on the successful realization of each VNF it contains, it is necessary to integrate the data of all VNFs to obtain the failure rate of the entire SFC (Lines 14-16). Only when this SFC does not fail can it bring an ideal return, based on which we will also make the backup selection and deployment of next round.
\begin{algorithm}\label{alg:failure}
       \caption{\textsc{GetFailureRate}($t$)}
       \SetKwBlock{Initialization}{Initialization}{end}
       \SetKwBlock{SFCPlacement}{SFCPlacement}{end}
       \SetKwBlock{OnlineLearning}{OnlineLearning}{end}
       \SetKwBlock{GetParameters}{GetParameters}{end}
       \KwIn{the current time slot $t$, the number of learning slots of failure \{$h_{i}(t)$\}, the average value of historical failure rate \{$\bar{v}_{i}(t)$\}}
       \KwOut{the failure rate of each VNF for next time slot \{$\Tilde{v}_{i}(t+1)$\}}
       \GetParameters{
          \ForEach{VNF $i \in \mathcal I$}{
            $h_{i}(t+1)=\begin{cases}
            h_{i}(t),             &\sum_{n\in \mathcal{N}}P_{i,n}(t)=0,\\
            h_{i}(t)+\sum_{n\in \mathcal{N}}P_{i,n}(t),           &\sum_{n\in \mathcal{N}}P_{i,n}(t)>0,
            \end{cases}$
            
            $\bar{v}_{i}(t+1)=\begin{cases}
            \bar{v}_{i}(t),          &\sum_{n\in \mathcal{N}}P_{i,n}(t)=0,\\
            \frac{h_{i}(t)\bar{v}_{i}(t)+V_{i}(t+1)}{h_{i}(t+1)},         &\sum_{n\in \mathcal{N}}P_{i,n}(t)>0.
            \end{cases}$
         }
        }
       \OnlineLearning{\ForEach{VNF $i \in \mathcal I$}{
             \If{$h_{i}(t+1)>0$}{
             $\Tilde{v}_{i}(t+1)={\bar{v}_{i}(t+1)+K\sqrt{\frac{3\log{t}}{2h_{i}(t+1)}}}$\;
             }
           }
       }
       \ForEach{SFC $f \in \mathcal F$}{
       $U_f(t)= \max_{i \in I_f} v_i(t)$\;
       }
\end{algorithm}

\subsection{Selection and Deployment}\label{subsec:Selection and deployment}
\subsubsection{Realization idea}
As shown in problem \eqref{eq:optimization_goal}, our goal is to maximize the time-average hit reward. Therefore, we need to choose and deploy backups of the most suitable SFCs at the edge with limited resources. In this case, our selection criterion is naturally the reward of each SFC.
\subsubsection{Algorithm implementation}
In Algorithm \ref{alg:VNFdeployment}, when the popularity, failure rate and minimum link delay of a certain SFC have been obtained by the previous algorithms, we can calculate the pre-reward of this SFC, expressed by $R_{f}^{'}$, according to \eqref{eq:reward} (Lines 1-3). Then we will choose the SFC that can bring the greatest reward, and deploy the VNFs it contains on determined edge servers according to the pre-deployment scheme with the least delay in Algorithm \ref{alg:Placement} (Lines 4-5), which will reduce the available resources of the corresponding edge servers (Lines 6-8). 
\begin{algorithm}\label{alg:VNFdeployment}
       \caption{\textsc{VNFdeployment}($t$)} 
       \KwIn{the current time slot $t$, the learned popularity of SFCs \{$\Tilde{q}_{f}(t)$\}, the lowest link latency consumed by each SFC \{$L_f(t)$\}}
       \KwOut{the placement decision of SFC backups $\{X_f(t)\}$}
       \ForEach{SFC $f\in\mathcal{F}$}{
       $R_{f}^{'}(t)=(\omega Q_{f}(t)-\mu L_{f}(t))*(1-U_f(t))$\;
       }
       Select SFC with the largest reward to place on the determined edge servers\;
       $X_f(t) \leftarrow 1$\;
       $P_{i,n}(t) \leftarrow 1$\;
        \ForEach{Edge Server $n \in \mathcal N $}{
          Update the resource of each server $M_n$\;
        }
\end{algorithm}

\section{Experimental Evaluation}\label{sec:experiment}
In this section, we evaluate our proposed solutions for SFC backup placement at the edge based on extensive simulations, which are conducted on a desktop with Intel(R) Core(TM) i7-9700 CPU @3.00GHz and 16GB RAM running Windows 10 OS.

\subsection{Experiment Settings}\label{Simulation Settings}
To facilitate experimental observation, we simulate an edge network with six servers. Each edge server has limited resource capacity which is assigned with a specific value in our experiment as presented in Table \ref{tab:Server set}. The link latency between any two servers is predetermined and known. In our experiments, we use a random function to generate a fixed delay to the link between any two servers as an example. Assuming that all users' service requests correspond to six SFCs which contain 15 VNFs in total and the specific resource requirements of VNFs are set according to Table \ref{tab:VNF set}. And Table \ref{tab:SFC set} presents the relationship between SFCs and VNFs.

\begin{table}[]
    \centering
    \footnotesize
    \caption{Resource limitation of edge servers.}
    \label{tab:Server set}
    \begin{tabular}{|c|c c c c c c|}
    \hline
    Server No.   &0 &1 &2 &3 &4 &5 \\
    \hline
    Available resource   &10 &8 &9 &12 &8 &11\\
    \hline
    \end{tabular}
\end{table}

\begin{table*}[]
    \centering
    \footnotesize
    \caption{VNF Setting.}
    \label{tab:VNF set}
    \begin{tabular}{|c|c c c c c c c c c c c c c c c|}
    \hline
    VNF No.   &0 &1 &2 &3 &4 &5 &6 &7 &8 &9 &10 &11 &12 &13 &14 \\
    \hline
    Resource requirement   &5 &4 &4 &8 &5 &3 &5 &8 &7 &5 &1 &4 &3 &3 &4\\
    \hline
    \end{tabular}
\end{table*}

\begin{table}[]
    \centering
    \footnotesize
    \caption{SFC Setting.}
    \label{tab:SFC set}
    \begin{tabular}{|c|c|}
    \hline
    $SFC0$   &$VNF3-VNF6-VNF9-VNF7-VNF4$ \\
    \hline
    $SFC1$   &$VNF9-VNF8-VNF1-VNF3$\\
    \hline
    $SFC2$   &$VNF3-VNF1-VNF6$ \\
    \hline
    $SFC3$   &$VNF10-VNF14-VNF1$ \\
    \hline
    $SFC4$   &$VNF1-VNF11-VNF13-VNF1-VNF4$\\
    \hline
    $SFC5$   &$VNF8-VNF1-VNF12-VNF10$\\
    \hline
    \end{tabular}
\end{table}

\subsection{Comparison Experimental Results}\label{comparative experiment}
Since our proposed \textit{RTSD} algorithm maximizes the time-average service hit reward by learning uncertain parameters in real time and comparing different plans in advance to adjust the selection and deployment strategy of SFC backups, to reflect the effectiveness of this adaptive scheme, we have designed two benchmark solutions for comparison. 

First of all, we verify the advantages of Algorithm \ref{alg:Placement} to show the effectiveness of our proposed SFC deployment scheme \textit{RTSD}. Specifically, in each time slot, we still estimate the popularity and failure rate of SFC using bandit learning method presented in Algorithm \ref{alg:popularity} and Algorithm \ref{alg:failure} so that we can calculate the total reward later. However, when deploying multiple VNFs contained in a specific SFC at the edge, instead of using the idea of Algorithm \ref{alg:Placement} inspired by \textit{Prim}, we place VNF involved in this SFC on the current server that can afford its resource requirements. Once the resource demand of the current VNF exceeds the remaining available resources of this server, we traverse down in the order of edge servers and select the next available server. According to the above scheme, we can get the corresponding delay of deploying each SFC in the edge environment. Combined with the popularity and failure rate learned before, we can calculate the service hit reward of all SFCs. We still choose the SFC with the largest reward value for actual deployment until the edge cannot tolerate any SFC. We refer to this benchmark solution as \textit{Bandit Scheme}.

Subsequently, we aim to verify the validity of the SFC real-time selection scheme with the essence being the idea of bandit learning as shown in Algorithm \ref{alg:popularity} and Algorithm \ref{alg:failure}. Therefore, instead of choosing SFC based on the reward value from the feedback, we randomly select an SFC that satisfies the edge resource constraint to be deployed at the edge. We put the VNFs involved in this SFC on qualified servers, where the resource requirement of each VNF does not exceed the existing available resource of the edge server. Repeating the above steps until the edge network cannot accommodate any SFC. For reference, we call this solution as \textit{Random Scheme}.

To compare our \textit{RTSD} algorithm with \textit{Bandit Scheme} and \textit{Random Scheme}, we analyze their performances from three aspects. Firstly, our SFC backup problem is from the users' point of view, which looks forward to bringing users the greatest reward. Therefore, the total service hit reward that SFCs can bring is a direct performance indicator that we need to observe. Secondly, our problem needs to meet the resource constraint of edge environment, but we hope to use the limited resources of the edge network as much as possible for cost-efficiency. In other words, the less unused resources remaining at the edge, the better performance the scheme. Therefore, the surplus of edge servers' resources is also one of our focuses. In addition, we also observed the total number of SFCs deployed at the edge in different schemes because deploying more SFC backups at the edge can satisfy users with more service requests. 

\begin{figure*}[htbp]
\centering
\subfigure[Hit reward]{\label{fig:reward}\includegraphics[width=0.29\textwidth]{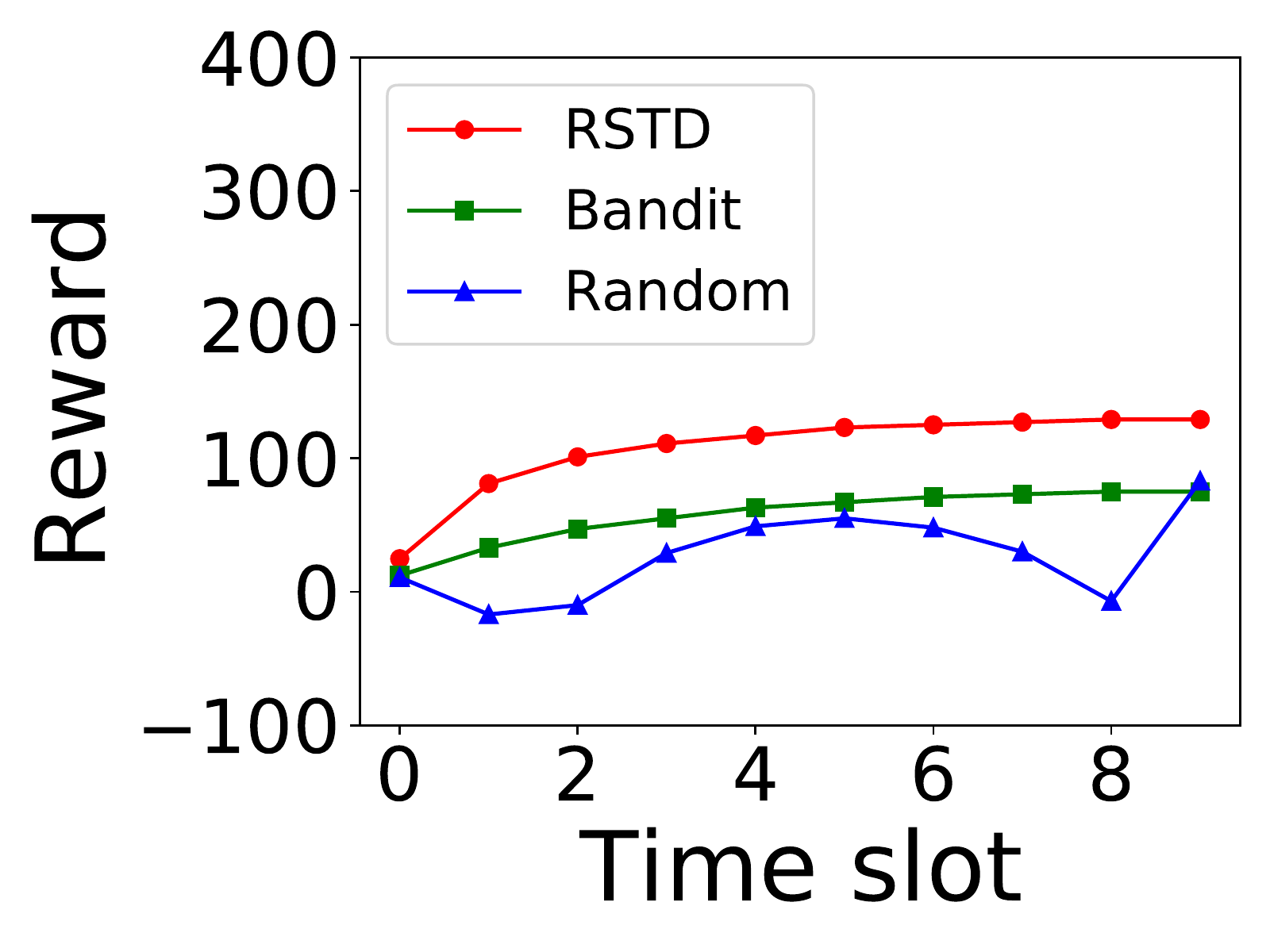}}
\centering
\subfigure[Remaining resource]{\label{fig:resource}\includegraphics[width=0.29\textwidth]{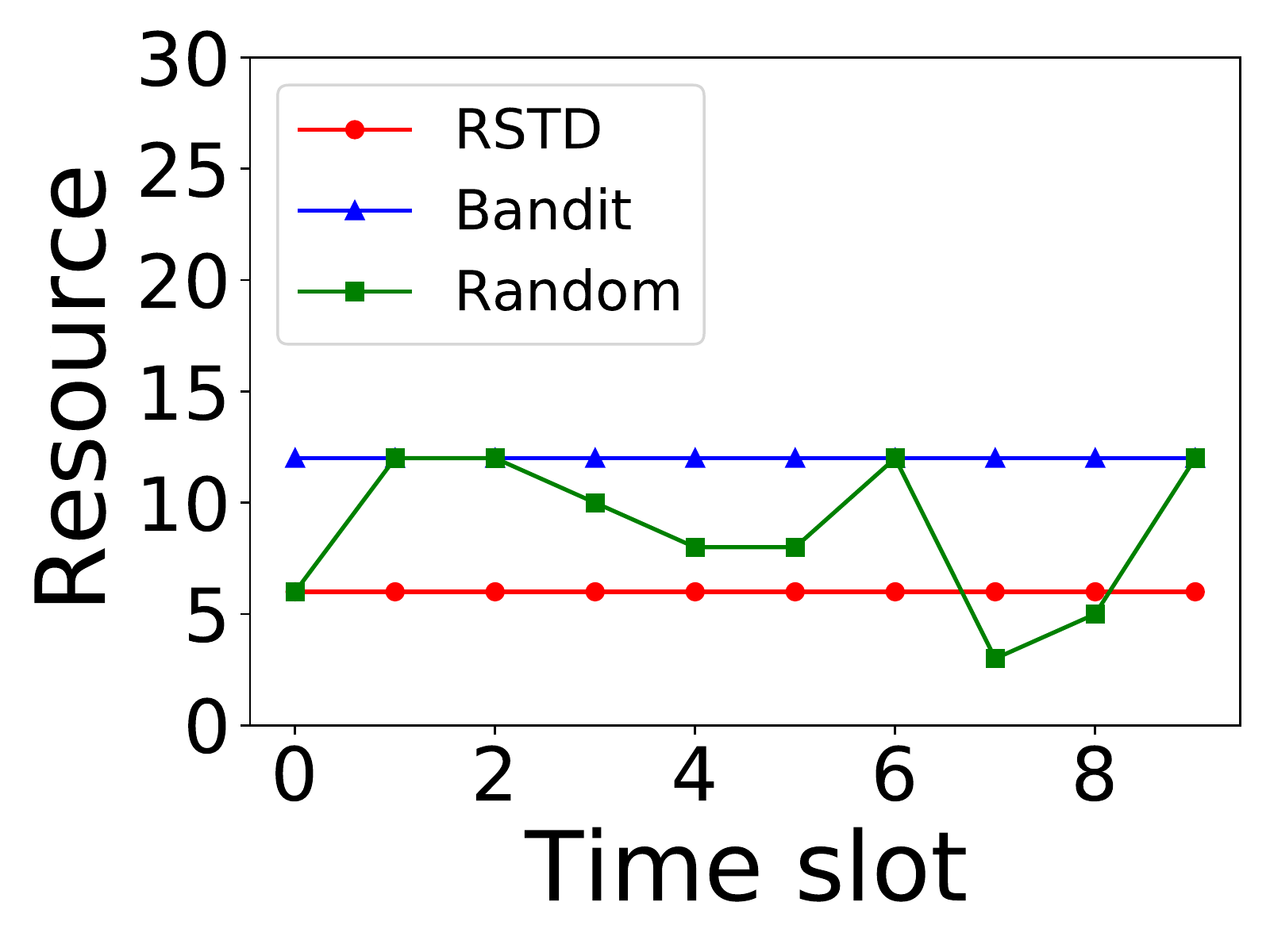}}
\centering
\subfigure[The number of SFC backups]{\label{fig:number}\includegraphics[width=0.29\textwidth]{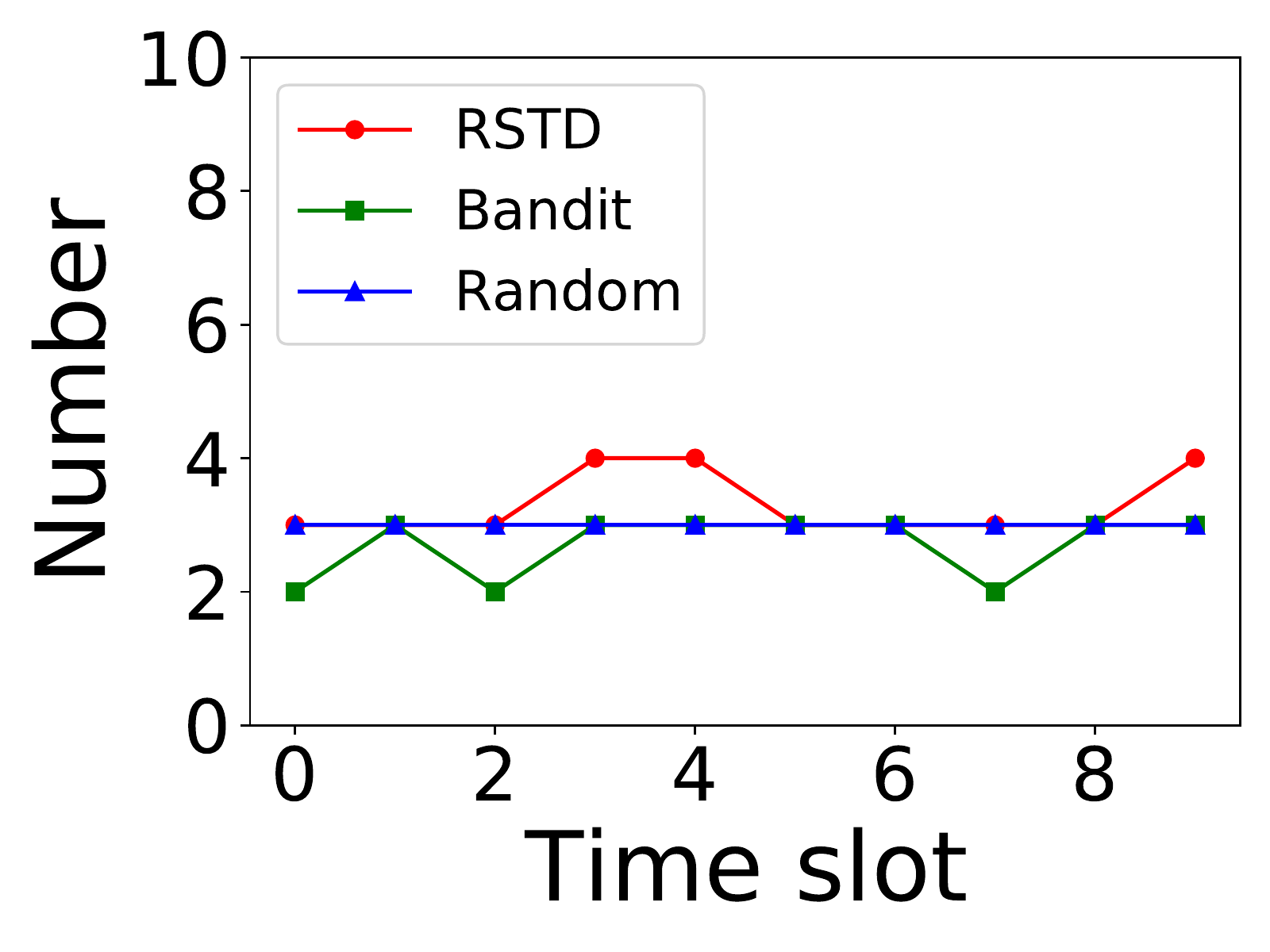}}
\caption{Comparison results between \textit{RTSD} and benchmark solutions.}
\label{fig:compare}
\end{figure*}

Fig. \ref{fig:compare} shows the comparison results, from which we can see that \textit{RTSD} solves the selection and deployment problem of SFC backup in the edge environment more effectively. Among them, we can observe that the service hit reward of \textit{Bandit Scheme} which uses bandit learning idea is higher than that of \textit{Random Scheme} as shown in Fig. \ref{fig:reward}, meaning that the Prim-inspired pre-deployment scheme applied in the \textit{RTSD} algorithm exerts better performance when deploying specific VNFs. Besides, \textit{RTSD} algorithm has shown the greatest advantage significantly, which further demonstrates that implementing online bandit bearning will indeed increase the benefits of our solution. And Fig. \ref{fig:resource} shows that compared with the benchmark solutions, our solution can achieve the least waste of resources at the edge. In other words, we can make the best use of the limited edge resources. Finally, the number of SFCs deployed on the edge is not significantly different in different schemes as shown in Fig. \ref{fig:number}, which is mainly limited by the available resources at the edge not changing too much in this simulation experiment.

\subsection{Performance Evaluation of \textit{RTSD}}
In this section, we focus on investigating the impacts of different parameters on the performance of the proposed \textit{RTSD} algorithm.

\begin{figure*}[htbp]
\centering
\subfigure[Hit reward]{\label{fig:User-reward}\includegraphics[width=0.29\textwidth]{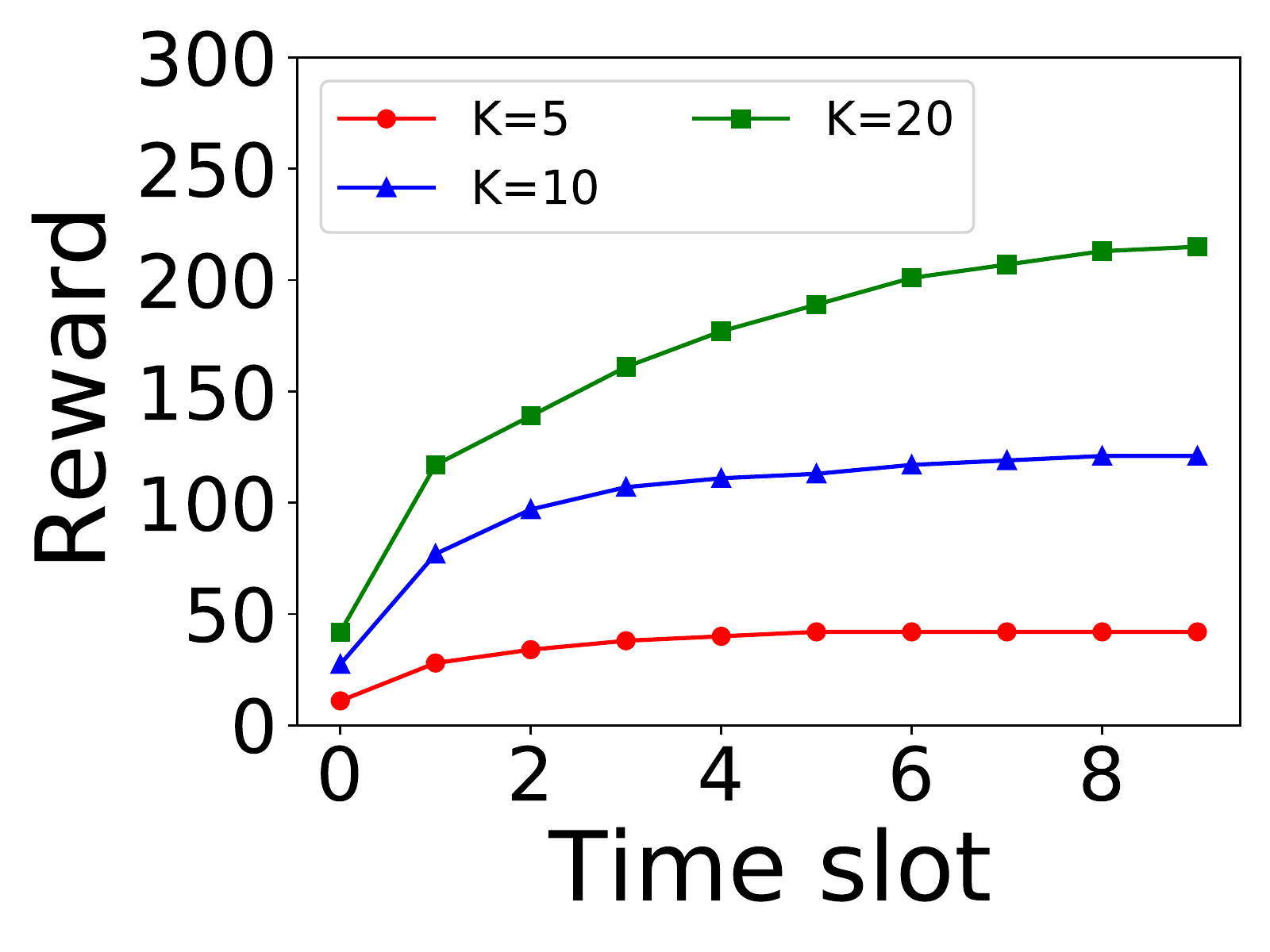}}
\centering
\subfigure[Remaining resource]{\label{fig:User-resource}\includegraphics[width=0.29\textwidth]{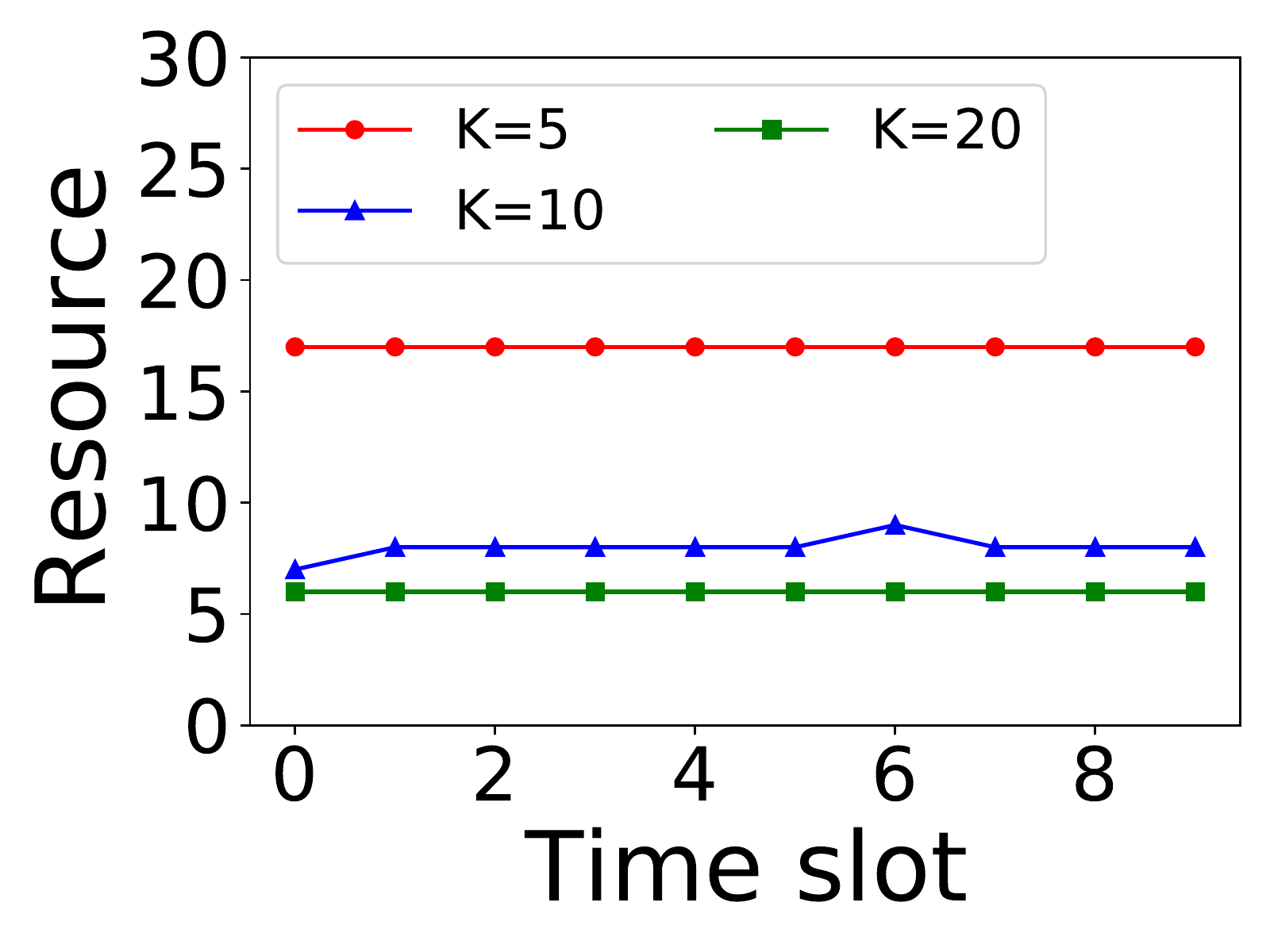}}
\centering
\subfigure[The number of SFC backups]{\label{fig:User-number}\includegraphics[width=0.29\textwidth]{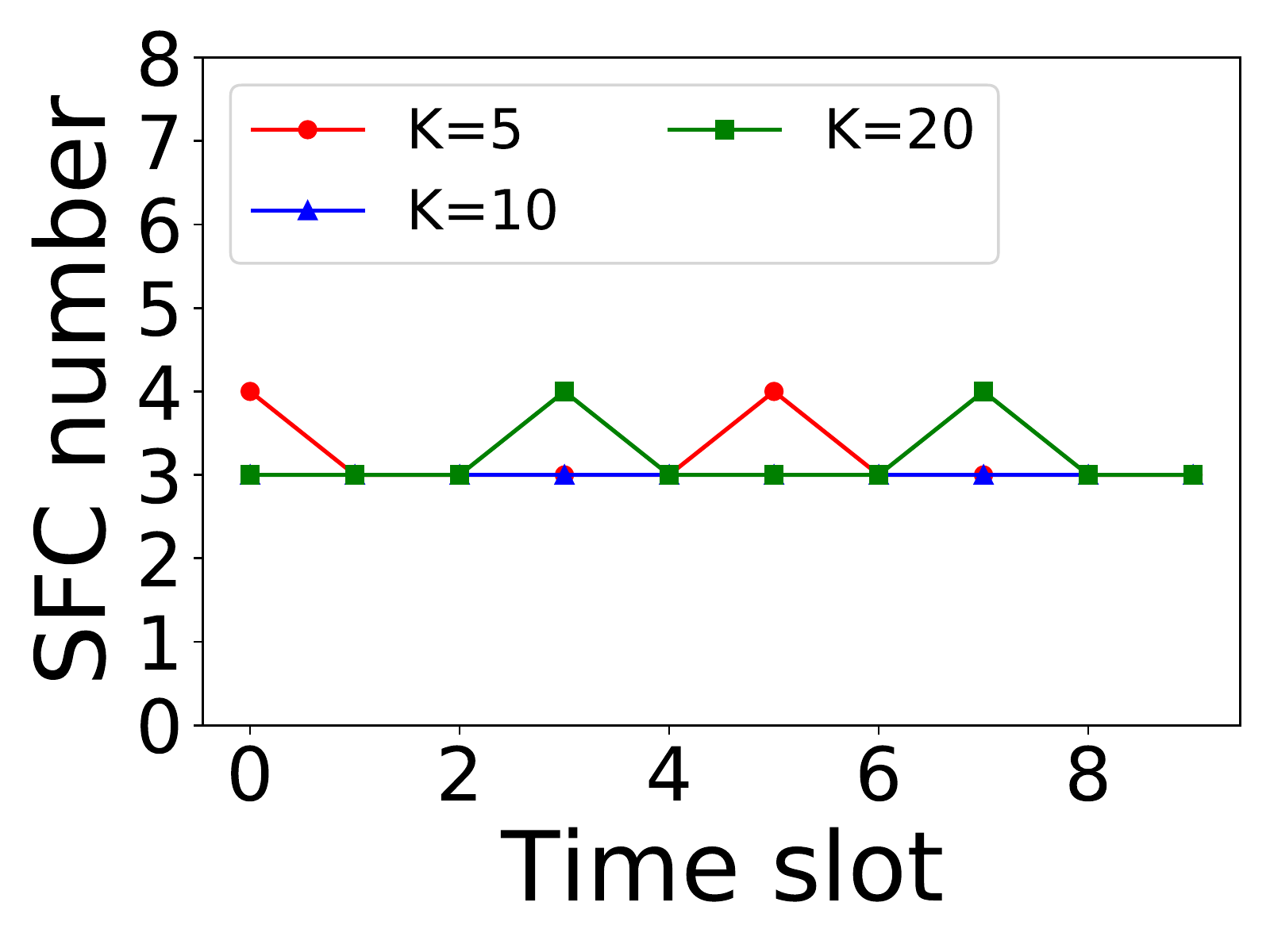}}
\caption{The effect of changing the number of users on \textit{RTSD}.}
\label{fig:change-User}
\end{figure*}

First, we observe the impact of end users' needs on our solution by changing the number of users in the simulation experiment. We set the number of users sending service requests as $5, 10$, and $15$, and then we get the corresponding changes in the service hit reward, remaining resources, and the number of deployed SFCs, which are shown in Fig. \ref{fig:change-User}. It can be seen that as the number of users increases, the request information we can refer to becomes more sufficient, which allows us to make full use of the limited resources at the edge and further obtain higher service hit reward by deploying more suitable SFCs at the edge. But the number of SFCs that can be deployed on edge servers will not change significantly due to the limitation of the available resources as we can see in Fig. \ref{fig:User-number}.

\begin{figure*}[htbp]
\centering
\subfigure[Hit reward]{\label{fig:Edge-reward}\includegraphics[width=0.29\textwidth]{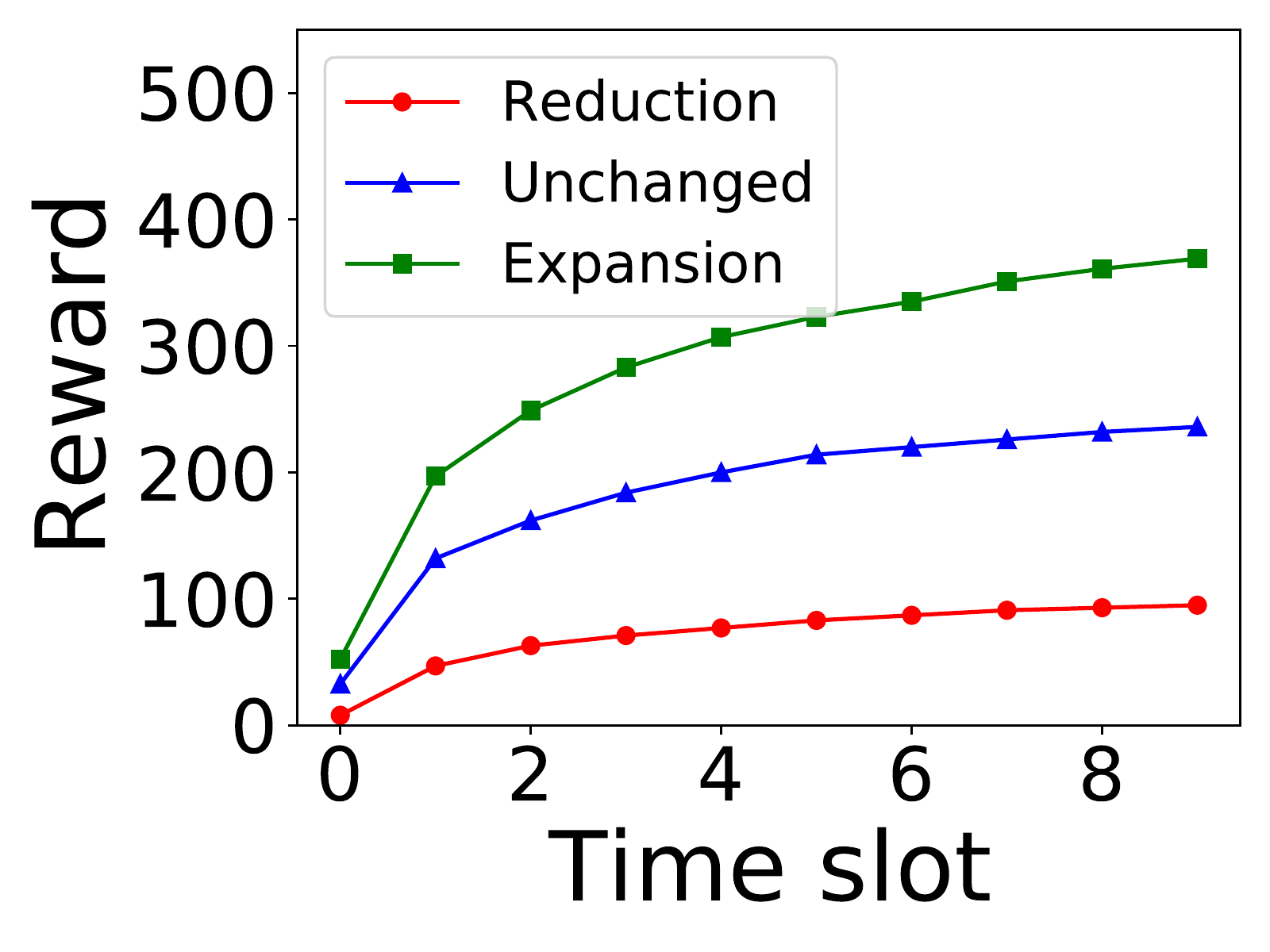}}
\centering
\subfigure[Remaining resource]{\label{fig:Edge-resource}\includegraphics[width=0.29\textwidth]{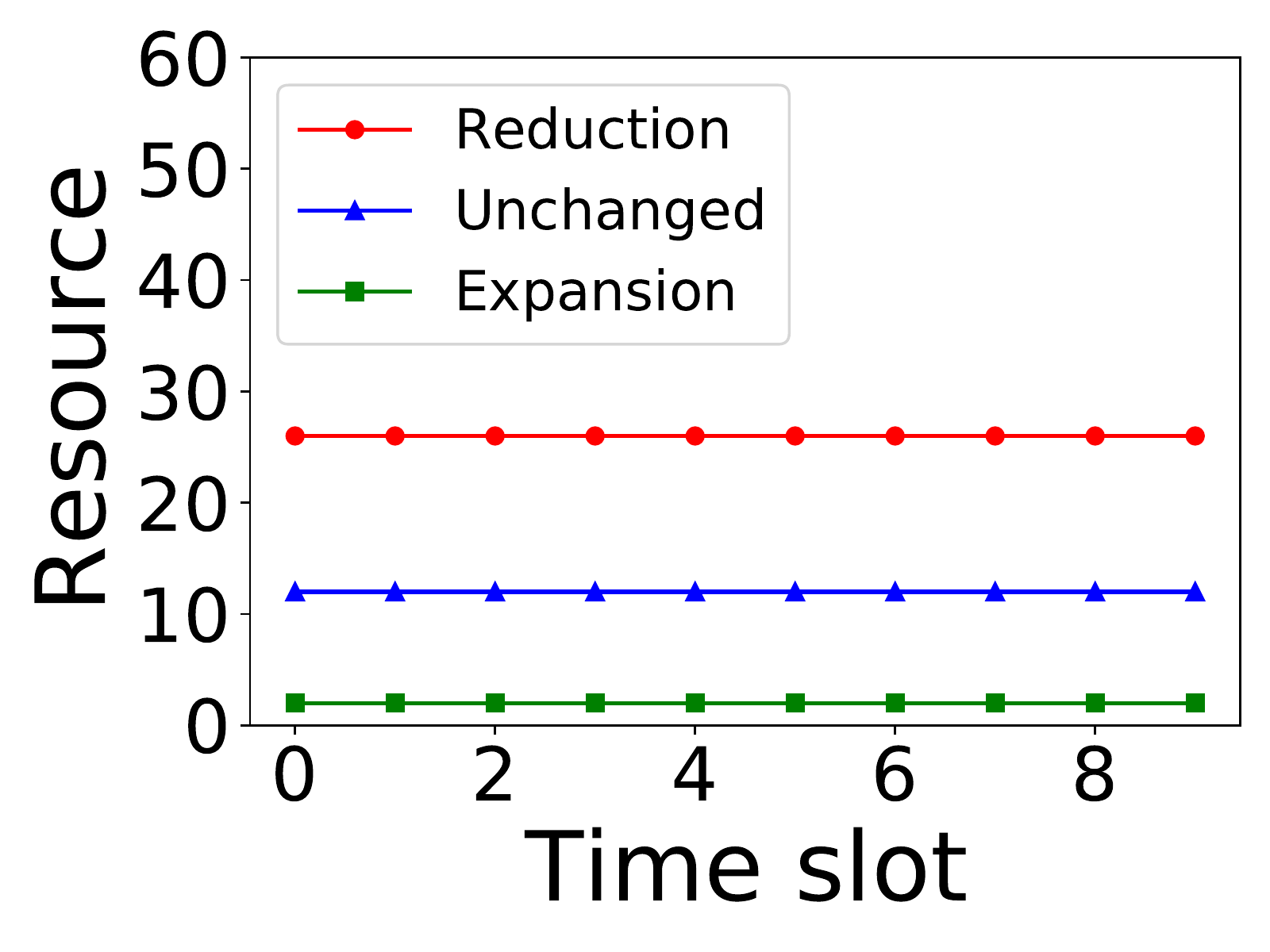}}
\centering
\subfigure[The number of SFC backups]{\label{fig:Edge-number}\includegraphics[width=0.29\textwidth]{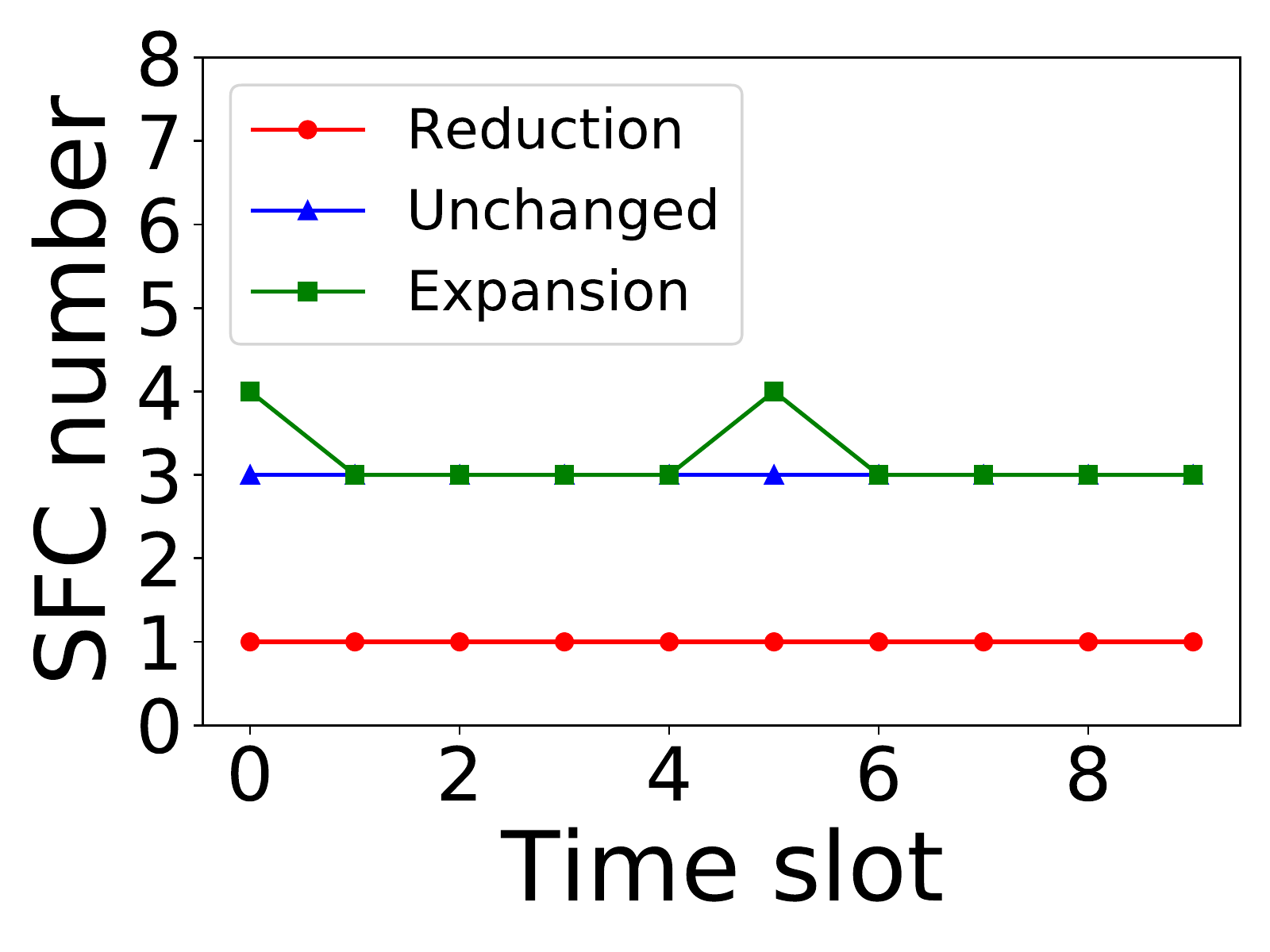}}
\caption{The effect of changing edge resources on \textit{RTSD}.}
\label{fig:change-Edge}
\end{figure*}

Subsequently, we observe the impact of changes in the edge environment on our \textit{RTSD} algorithm. By reducing the available resources of the edge server by $50\%$ and increasing it by $50\%$, Fig. \ref{fig:change-Edge} illustrates that when the servers' available resources increase, we can deploy more backups at the edge, which will bring less waste of resources and more service reward.

\section{Conclusion}\label{sec:conclusion}
In this paper, we solve the problem of deploying SFC backups at the edge. To cope with the shortcomings of current research, we conduct edge backup of SFCs to provide popular services with the lowest latency considering from the perspectives of both the end users and edge system.
To that aim, we propose a \textit{Real-Time Selection and Deployment} (\textit{RTSD}) algorithm. We first take advantage of the online bandit learning method to deal with the uncertainty of SFC popularity and failure rate dynamically changing with time. Next, inspired by the \textit{Prim} method and combined it with the greedy strategy, we find the optimal deployment plan with the minimum latency for a specific SFC. Backup hit rewards are calculated by integrating the results of the two steps, based on which we can select popular SFC backups and properly place the corresponding VNFs contained in them on the edge network. The deployment result of the current time slot will be used as feedback information to optimize the subsequent SFC backup operation.
Results of comparative experiments demonstrate the superiority of our proposed schemes.

\ifCLASSOPTIONcompsoc
  \section*{Acknowledgments}
\else
  \section*{Acknowledgment}
\fi

This work has been partially supported by the National Key Research and Development Program of China under Grant 2019YFB2102600, the National Natural Science Foundation of China (NSFC) under Grants 61971269, 61832012, and 61771289, and the US NSF under Grant CNS-2105004.



%
%



\par\noindent 
\parbox[t]{\linewidth}{
\noindent\parpic{\includegraphics[height=1.5in,width=1in,clip,keepaspectratio]{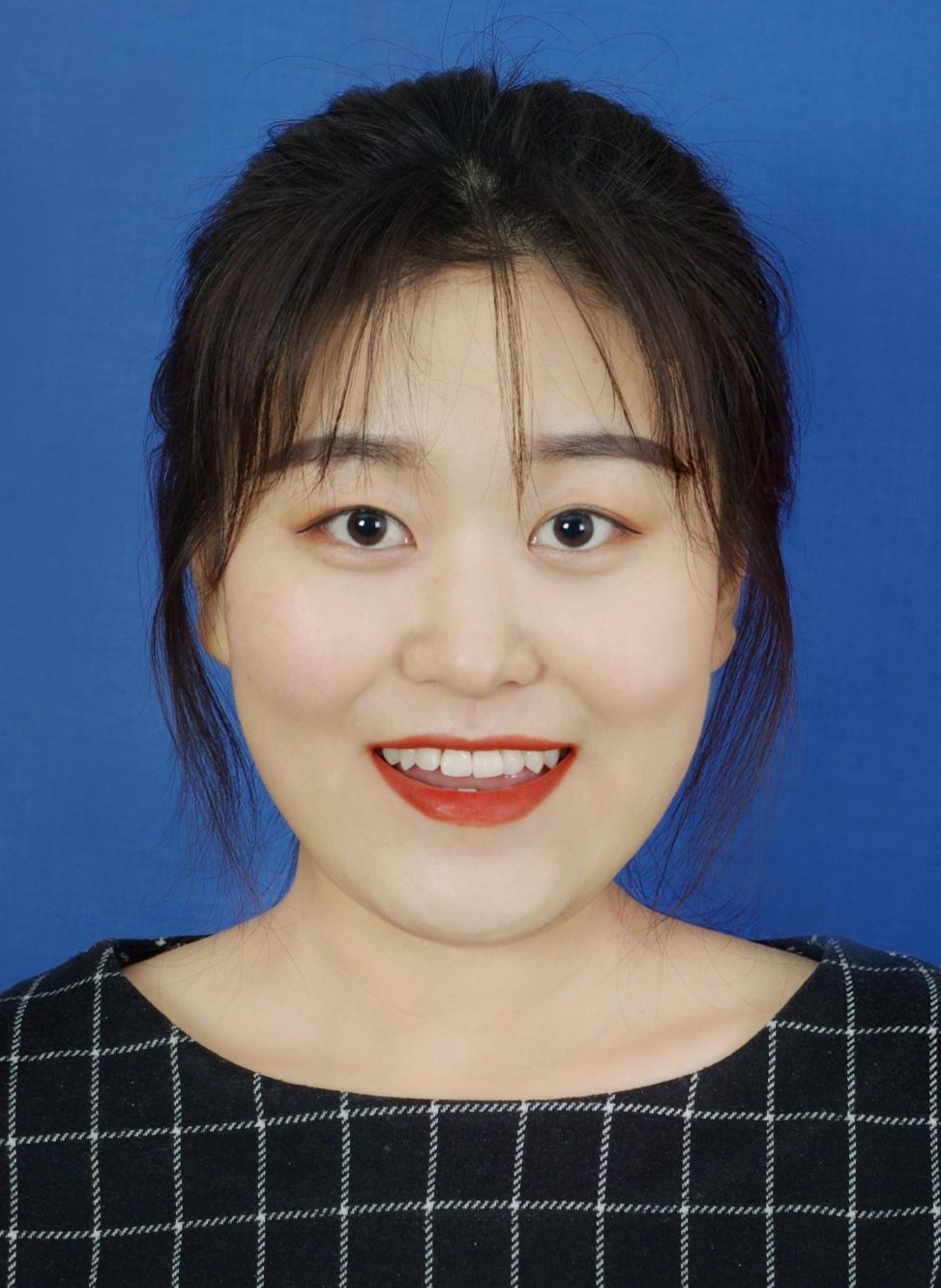}}
\noindent {\bf Chen Wang}\
received her BS degrees in Computer Science from China University of Mining and Technology at Xuzhou, China, in 2010. She is currently pursuing the MS degree with the School of Computer Science and Technology, Shandong University. Her research interests include edge computing and reinforcement learning.}
\vspace{4\baselineskip}

\par\noindent 
\parbox[t]{\linewidth}{
\noindent\parpic{\includegraphics[height=1.5in,width=1in,clip,keepaspectratio]{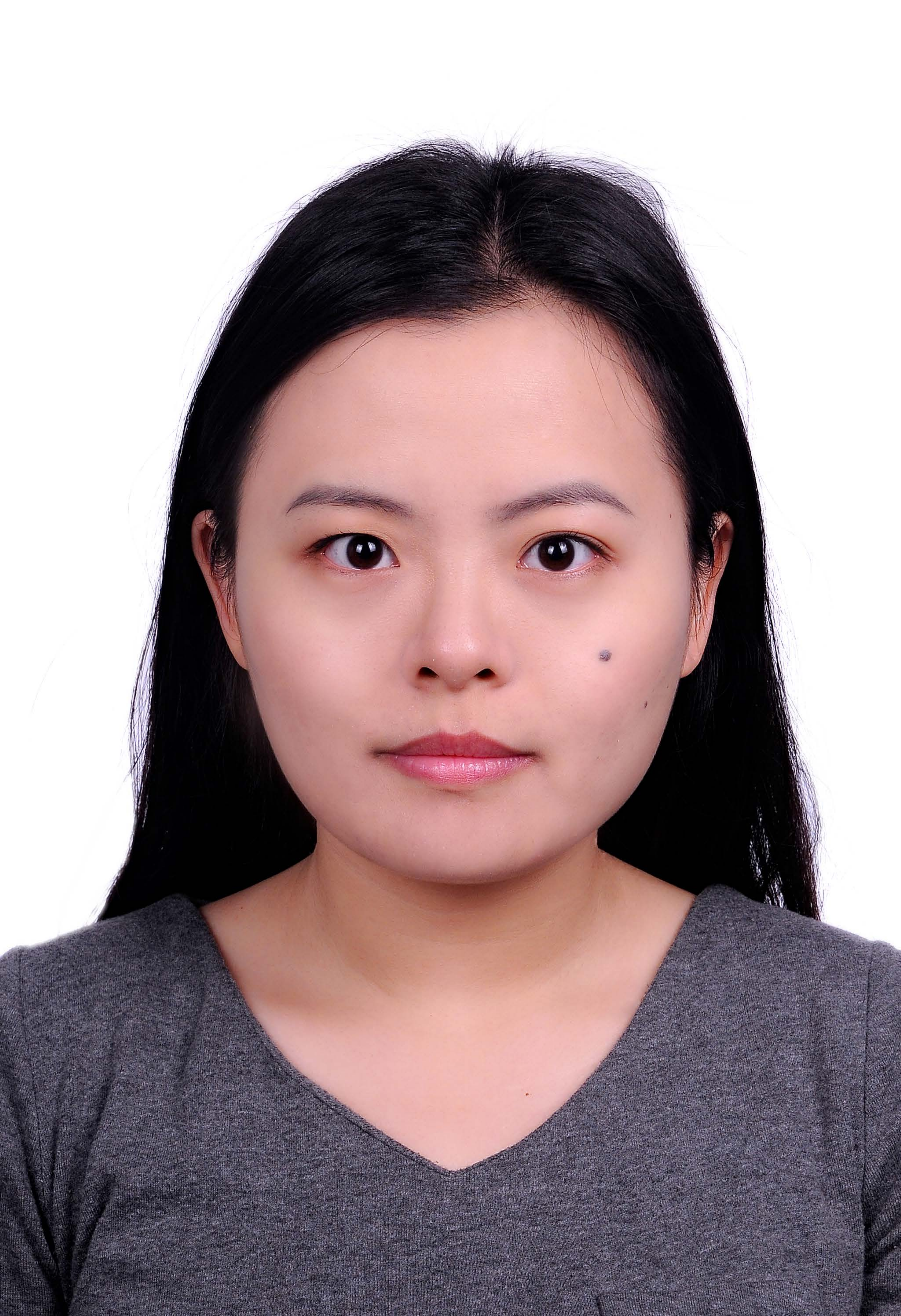}}
\noindent {\bf Qin Hu}\
received her Ph.D. degree in Computer Science from the George Washington University in 2019. She is currently an Assistant Professor with the Department of Computer and Information Science, Indiana University-Purdue University Indianapolis (IUPUI). Her research interests include wireless and mobile security, edge computing, blockchain, and crowdsourcing/crowdsensing.}
\vspace{4\baselineskip}

\par\noindent 
\parbox[t]{\linewidth}{
\noindent\parpic{\includegraphics[height=1.5in,width=1in,clip,keepaspectratio]{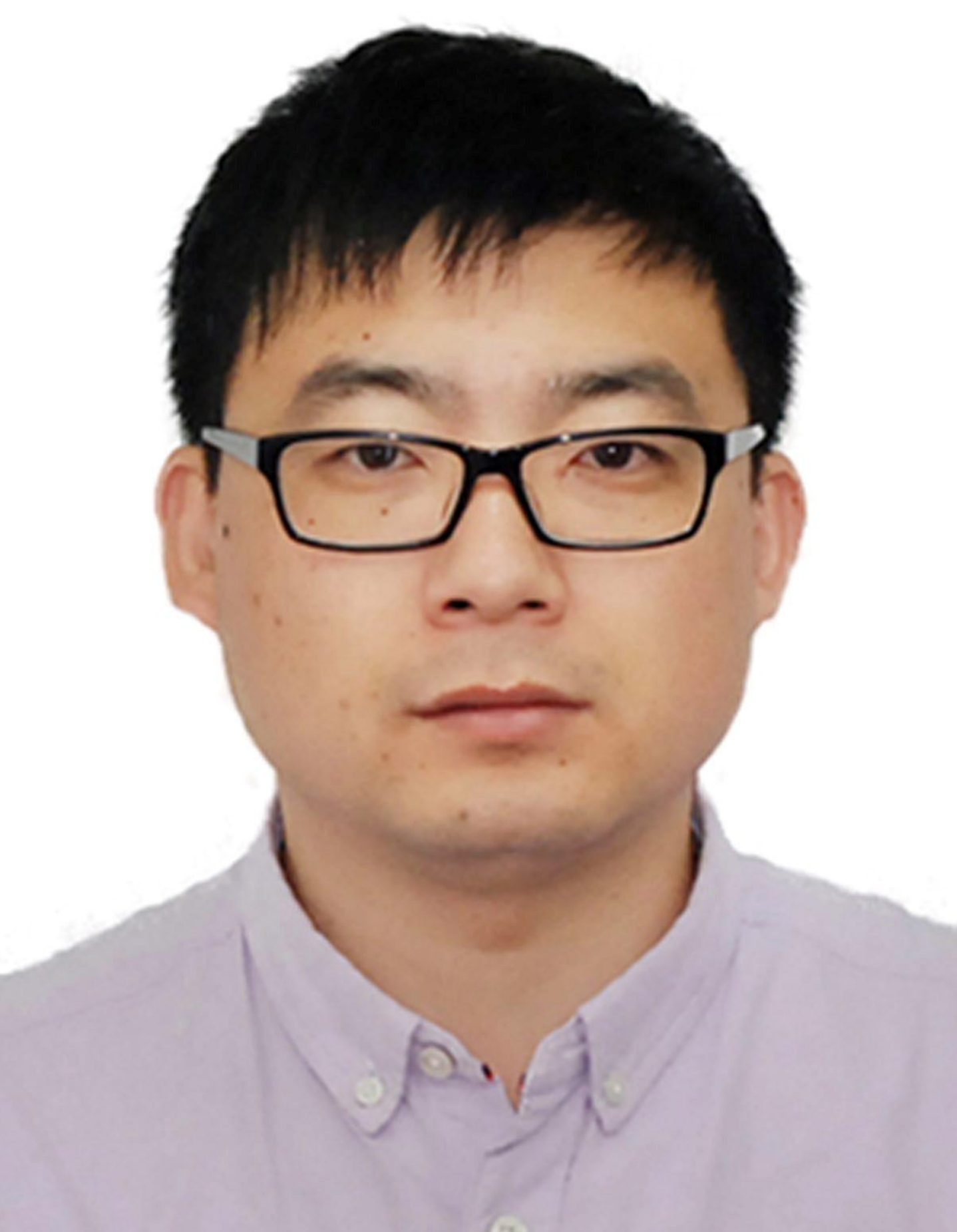}}
\noindent {\bf Dongxiao Yu}\
received his BS degree in 2006 from the School of Mathematics, Shandong University and the PhD degree in 2014 from the Department of Computer Science, The University of Hong Kong. He became an Associate Professor in the School of Computer Science and Technology, Huazhong University of Science and Technology, in 2016. Currently, he is a professor in the School of Computer Science and Technology, Shandong University. His research interests include wireless networks, distributed computing, and data mining.}
\vspace{4\baselineskip}

\par\noindent 
\parbox[t]{\linewidth}{
\noindent\parpic{\includegraphics[height=1.5in,width=1in,clip,keepaspectratio]{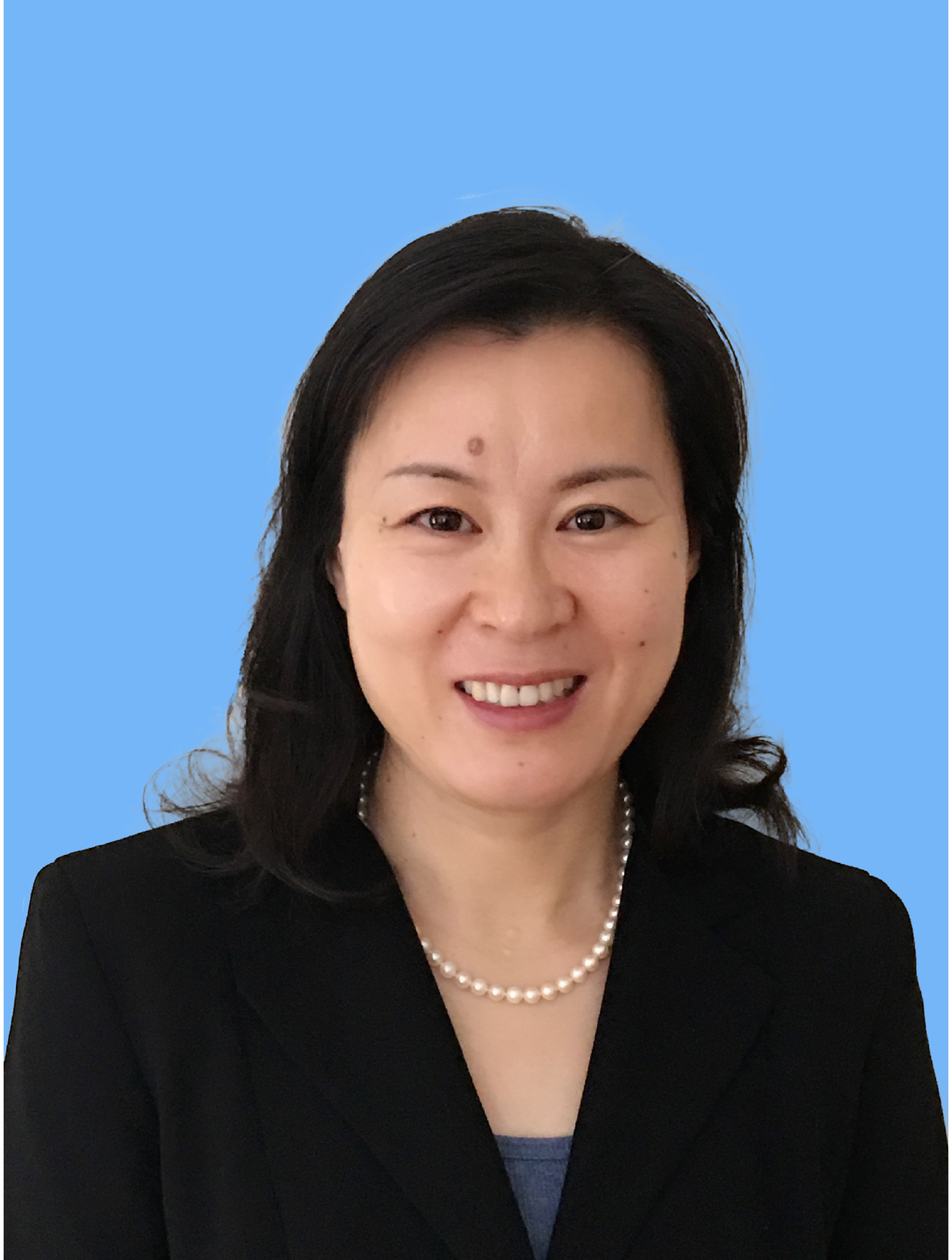}}
\noindent {\bf Xiuzhen Cheng}\
received her M.S. and Ph.D. degrees in computer science from the University of Minnesota Twin Cities in 2000 and 2002. She is currently a professor in the School of Computer Science and Technology, Shandong University, China. Her current research interests focus on privacy-aware
computing, wireless and mobile security, dynamic spectrum access, mobile handset networking systems (mobile health and safety), cognitive radio networks, and algorithm design and analysis. She has served on the Editorial Boards of several technical publications and the Technical Program Committees of various professional conferences/workshops. She has also chaired several international conferences. She worked as a program director for the U.S. National Science Foundation (NSF) from April to October 2006 (full time), and from April 2008 to May 2010 (part time). She published more than 300 peer-reviewed papers.}
\vspace{4\baselineskip}

\end{document}